\documentclass{article}

 \usepackage[margin=2cm]{geometry}

 \usepackage{graphics,arydshln}
\usepackage{graphicx} 
\usepackage{amssymb, natbib}
\usepackage{amsfonts, amsmath, amssymb, marvosym,colonequals}
\usepackage{algorithm,multirow}
\usepackage{bm}
\usepackage{geometry}
\usepackage{lscape}
\usepackage{mathtools}
\usepackage{caption}
\usepackage{theorem}
\usepackage{placeins}
\usepackage{enumerate}
\usepackage{epsfig}
\usepackage[all]{xy}
\usepackage{varwidth}

\usepackage{color,soul}

\newcommand{\be}{\begin{equation}}
\newcommand{\ee}{\end{equation}} 
\newcommand{\beq}{\begin{equation*}}
\newcommand{\eeq}{\end{equation*}}
\newcommand{\vecx}{\mathbf{x}}
\newcommand{\vecX}{\mathbf{X}}

\newcommand{\vecvarthet}{\mbox{\boldmath$\vartheta$}}

\newcommand{\ident}{\mathbf{I}}

\newcommand{\vecmu}{\mbox{\boldmath$\mu$}}

\newcommand{\vecalpha}{\mbox{\boldmath$\alpha$}}

\newcommand{\matsig}{\mathbf{\Sigma}}

\newtheorem{Theorem}{Theorem}[section]
\newtheorem{Prop}[Theorem]{Proposition}

\theorembodyfont{\upshape}

\date{}
\begin{document}

%\begin{frontmatter}

\hyphenation{mcnicholas}

\title{Clustering, Classification, Discriminant Analysis, and Dimension Reduction via Generalized Hyperbolic Mixtures}
\author{Katherine Morris\thanks{Department of Mathematics \& Statistics, University of Guelph, Ontario, Canada, N1G~2W1.} \qquad Paul D.\ McNicholas\thanks{Department of Mathematics \& Statistics, McMaster University, 1280 Main St.\ W., Hamilton, Ontario, Canada L8S~4L8. Email: {\tt mcnicholas@math.mcmaster.ca}}} %\thanks{Department of Mathematics \& Statistics, University of Guelph, 

\maketitle

\begin{abstract}
A method for dimension reduction with clustering, classification, or discriminant analysis is introduced. This mixture model-based approach is based on fitting generalized hyperbolic mixtures on a reduced subspace within the paradigm of model-based clustering, classification, or discriminant analysis. A reduced subspace of the data is derived by considering the extent to which group means and group covariances vary. The members of the subspace arise through linear combinations of the original data, and are ordered by importance via the associated eigenvalues. The observations can be projected onto the subspace, resulting in a set of variables that captures most of the clustering information available. The use of generalized hyperbolic mixtures gives a robust framework capable of dealing with skewed clusters. Although dimension reduction is increasingly in demand across many application areas, the authors are most familiar with biological applications and so two of the five real data examples are within that sphere. Simulated data are also used for illustration. The approach introduced herein can be considered the most general such approach available, and so we compare results to three special and limiting cases. Comparisons with several well established techniques illustrate its promising performance.\\

\end{abstract}

\section{Introduction}
A method for estimating a projection subspace basis derived from the fit of a generalized hyperbolic mixture (HMMDR) is introduced within the paradigms of model-based clustering, classification, and discriminant analysis. This is the most general case of work in this direction over the last few years, starting with an analogous approach based on Gaussian mixtures \citep[GMMDR;][]{scrucca10}. 

Many dimension reduction methods summarize the information available through a reduced combination of the original variables. However, in terms of visualization, they do not always provide adequate information on the potential structure of the data at hand. The method proposed herein addresses this issue by revealing the underlying data clusters. At the same time, using heavy-tailed distributions, such as the generalized hyperbolic distribution, to model data can be advantageous because they assign appropriate weights to more extreme points \citep{mcneil05}. The goal is to estimate a subspace that captures most of the clustering structure contained in the data. At the core of the method lies the sliced inverse regression (SIR) work of \cite{li91,li00}, which reduces data dimensionality by considering the variation in group means to identify the subspace. \cite{scrucca10} extended the SIR ideas to also include variation of group covariances. The members of the subspace arise through linear combinations of the original data, and are ordered by importance via their associated eigenvalues. The original observations in the data can be projected onto the subspace, resulting in a set of variables that captures most of the clustering information available.

The remainder of the paper is outlined as follows. Section~\ref{sec:background} presents the background material. We then outline our dimension reduction method for selecting a reduced combination of the variables while retaining most of the clustering information contained within the data (Section~\ref{sec:methodology}). In Section~\ref{sec:applications}, the algorithm is applied to simulated and real data sets and the performance of our method is compared with its Gaussian and non-Gaussian analogues as well as with other subspace clustering techniques. Section~\ref{sec:conclusion} provides conclusions and suggestions for future work. Note that all computational work herein was carried out using {\sf R} \citep{R12}.

\section{Background}\label{sec:background}

\subsection{Finite Mixture Models}
Modern data sets used in many practical applications have grown in size and complexity, compelling the use of clustering and classification algorithms based on probability models. The model-based approach assumes that data are generated by a finite mixture of probability distributions. A $p$-dimensional random vector $\vecX$ is said to arise from a parametric finite mixture distribution if its density is a convex set of probability densities, i.e.,  
\begin{equation*}
f(\vecx~|~\vecvarthet)=\sum_{g=1}^G\pi_g\, f_g(\vecx~|~\bm \theta_g),
\end{equation*}
where $G$ is the number of components, $\pi_g$ are mixing proportions, so that $\sum_{g=1}^G\pi_g=1$ and $\pi_g>0$, and $\bm\vartheta=(\pi_1, \ldots, \pi_G, \bm \theta_1,\ldots, \bm \theta_G)$ is the parameter vector. The $f_g(\vecx| \bm\theta_g)$ are called component densities and $f(\vecx |\vecvarthet)$ is formally referred to as a $G$-component parametric finite mixture distribution. The use of mixture models in clustering applications can be traced back a half-century to an application of Gaussian mixture models \citep{wolfe63}. Gaussian mixture model-based approaches have been very popular due to their mathematical tractability, and until recently, they dominated literature in the field. Extensive details on finite mixture models are given by \cite{everitt81}, \cite{mclachlan88}, and \cite{mclachlan00b}.

In the past several years, non-Gaussian approaches to model-based clustering, classification, and discriminant analysis have flourished. This includes work on mixtures of multivariate $t$-distributions \citep{peel00, greselin10a, andrews11b, steane12,andrews12,mcnicholas13}, shifted asymmetric Laplace distributions \citep{franczak12}, skew-normal distributions \citep{lin10}, skew $t$-distributions \citep{vrbik12,vrbik14,lee13,murray14b,murray14a}, and variance-gamma distributions \citep{smcnicholas13}. Mixtures of generalized hyperbolic distributions \citep{browne15} are particularly relevant to work described herein. While it is not feasible to provide an exhaustive listing here, suffice it to say that the breadth of research on non-Gaussian model-based clustering and classification is becoming as rich as that of its Gaussian precursor.

Generalized hyperbolic distributions were introduced by \cite{barndorff77a} and used to model eolian sand deposits, i.e., sand deposits arising from the action of wind. The name of the distribution was derived from the fact that its log-density has the shape of a hyperbola. %(cf. log-density of the normal distribution is a parabola).
Properties of generalized hyperbolic densities were discussed in \cite{barndorff77b} and \cite{blaesild78} and, more recently, mixtures of these distributions appear in \cite{mcneil05} and \cite{hardle11}. Generalized hyperbolic distributions can effectively model extreme values, making them very useful in the context of financial and risk management applications, where the normal distribution does not offer a good description of reality. The multivariate generalized hyperbolic family is extremely flexible and contains many special and limiting cases, such as the inverse Gaussian, Laplace, and skew-$t$ distributions.

\subsection{Generalized Hyperbolic Mixtures}
\cite{browne15} propose a multivariate generalized hyperbolic mixture model (HMM),
\be
f(\vecx~|~\vecvarthet)=\sum_{g=1}^G\pi_g\, f_{\text{h}}(\vecx~|~\lambda_g, \omega_g, \bm\mu_g, \bm\Sigma_g, \bm\alpha_g)\,,
\label{eq:hmm}
\ee
where $\pi_g>0$, with $\sum_{g=1}^G\pi_g=1$, are the mixing proportions and the $g$th component density is
\begin{eqnarray}
f_{\text{h}}(\vecx~|~\lambda_g, \omega_g, \bm\mu_g, \bm\Sigma_g, \bm\alpha_g)&=& \displaystyle \left[\frac{\omega_g+\delta(\vecx, \vecmu_g | \matsig_g)}{\omega_g + \vecalpha_g^{\top}\matsig_g^{-1} \bm\alpha_g}\right]^{(\lambda_g -p/2)/2} \notag \\
&\times& \displaystyle \frac{K_{\lambda_g-p/2}\left(\sqrt{\left[\omega_g + \vecalpha_g^{\top}\matsig_g^{-1} \bm\alpha_g)(\omega_g+\delta(\vecx, \vecmu_g | \matsig_g)\right]}\right)}{(2\pi)^{p/2} \left | \bm\Sigma_g \right |^{1/2} K_{\lambda_g}(\omega_g) \exp(-(\vecx-\vecmu_g)^{\top}\bm\Sigma_g^{-1} \bm\alpha_g)}\,,
\label{eq:hdensity}
\end{eqnarray}
with index parameter $\lambda_g$, concentration parameter $\bm\omega_g$, skewness parameter $\bm\alpha_g$, location $\bm\mu_g$, and scale matrix $\bm\Sigma_g$.  Here, $\delta(\vecx, \vecmu_g~|~\matsig_g)=(\vecx-\vecmu_g)^{\top}\matsig_g^{-1}(\vecx-\vecmu_g)$ is the squared Mahalanobis distance between $\vecx$ and $\vecmu_g$ and $K_{\lambda_g}$ denotes the modified Bessel function of the third kind with index $\lambda_g$. 

The evaluation of modified Bessel functions in the density (\ref{eq:hdensity}) sometimes leads to numerical overflow or underflow. To avoid these issues, we use asymptotic expansions from \cite{abramowitz72}, i.e., for large $x$ or $\lambda$,
\begin{equation*}
K_\lambda(\lambda x)=\sqrt{\frac{\pi}{2\lambda}} \frac{\exp\{-\lambda\rho\}}{(1+x^2)^{1/4}}\left[ 1+ \sum_{k=1}^{\infty} (-1)^k \frac{u_k(\tau)}{\lambda^k}\right],
\end{equation*}
where $$\rho = \sqrt{1+x^2}+\ln\left(\frac{x}{1+\sqrt{1+x^2}}\right),$$ and $u_k(\tau)$ is the Debye polynomial represented by $u_0(\tau)=1$ and $$u_{k+1}(\tau)=\frac{1}{2}\tau^2 (1-\tau^2) u'_k(\tau)+\frac{1}{8}\int_0^{\tau} (1-5s^2) u_k(s) ds,$$ for $\tau=1/\sqrt{1+x^2}$ and $k=1,2,\ldots$
%\begin{equation*}
%u_0(\tau)=1\ \mathrm{and}\ u_{k+1}(\tau)=\frac{1}{2}\tau^2 (1-\tau^2) u'_k(\tau)+\frac{1}{8}\int_0^{\tau} (1-5s^2) u_k(s) ds,\ \mathrm{for}\ k=1, \ldots
%\end{equation*}
%with $\tau=1/\sqrt{1+x^2}$.

The parametrization in (\ref{eq:hdensity}) is one of several available for generalized hyperbolic distributions \citep[cf.][]{mcneil05}. In this case, the $p$-dimensional random vector $\vecX$ is generated by combining a generalized inverse Gaussian (GIG) random variable $Y$ with a latent multivariate Gaussian random variable $\bm U \sim \mathcal{N}(\bm 0, \matsig)$. Note that the density of $Y\sim\text{GIG}(\omega, \eta, \lambda)$ is
\beq
h(y~|~\omega, \eta, \lambda) = \frac{(y/\eta)^{\lambda-1}}{2\eta K_{\lambda}(\omega)} \exp\left\{ -\frac{\omega}{2}\left(\frac{y}{\eta} + \frac{\eta}{y}\right) \right\}.
\eeq
We fix $\eta=1$ and use the relationship $\vecx = \vecmu + Y\vecalpha +\sqrt{Y}\bm U$. Full details on the derivation of this parametrization and its use in parameter estimation are given by \cite{browne15}.

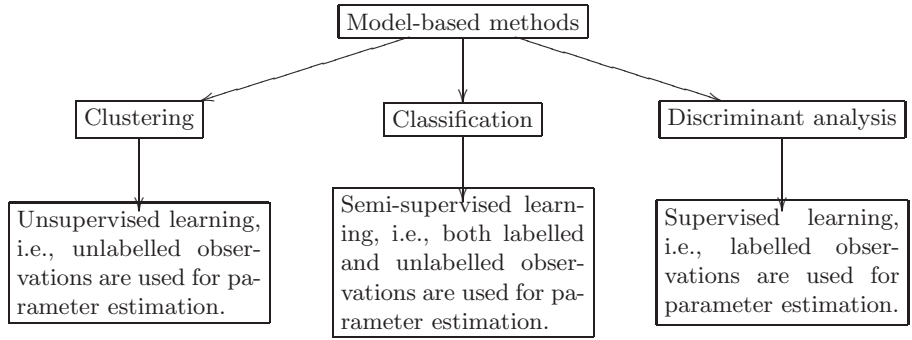
\begin{figure}{\small
\begin{displaymath}
\xymatrix {&*+[F]{\text{Model-based methods}}\ar[dl]\ar[d]\ar[dr]&\\
		*+[F]{\text{Clustering}}\ar[d]&*+[F]{\text{Classification}}\ar[d]&*+[F]{\text{Discriminant analysis}}\ar[d]\\
		*+[F]{\text{\begin{varwidth}{10em}Unsupervised learning, i.e., unlabelled observations are used for parameter estimation.\end{varwidth}}}&*+[F]{\text{\begin{varwidth}{10em}Semi-supervised learning, i.e., both labelled and unlabelled observations are used for parameter estimation.\end{varwidth}}}&*+[F]{\text{\begin{varwidth}{10em}Supervised learning, i.e., labelled observations are used for parameter estimation.\end{varwidth}}}}
\end{displaymath}}
\caption{Diagram showing the relationship between the model-based learning paradigms discussed.}
\label{fig:model-based}
\end{figure}

In the following sections, we discuss using the generalized hyperbolic distribution for model-based methods in the context of unsupervised (clustering), semi-supervised (classification) and supervised (discriminant analysis) learning. Figure \ref{fig:model-based} shows the relationship between these learning approaches.

\subsection{Model-Based Clustering}

Consider a clustering scenario in which none of the observations have known component membership, i.e., where the observations are unlabelled. The generalized hyperbolic model-based clustering likelihood is
\begin{equation}
\mathcal{L}(\vecvarthet~|~\vecx)=\prod_{i=1}^n\sum_{g=1}^G\pi_g f_{\text{h}}(\vecx_i~|~\lambda_g, \omega_g, \bm\mu_g, \bm\Sigma_g, \bm\alpha_g).
\label{eq:likecluster}
\end{equation}
To facilitate discussion of parameter estimation, introduce $Z_{ig}$ to denote component membership labels, so that $z_{ig}=1$ if observation $\vecx_i$ belongs to component~$g$ and $z_{ig}=0$ otherwise.  

Parameter estimation for generalized hyperbolic mixtures is carried out using the expectation-maximization (EM) algorithm \citep{baum70,orchard72,sundberg74,dempster77}. The EM algorithm is an iterative procedure for finding maximum likelihood estimates when data are incomplete or are treated as being incomplete. The EM algorithm is based on the complete-data log-likelihood \eqref{eq:loglikecluster}, where the complete-data comprise the observed ${\vecx}_i$, the missing $z_{ig}$, and the latent $y_{ig}$, for $i=1,\ldots,n$ and $g=1,\ldots,G$. Our complete-data log-likelihood can be written
\begin{eqnarray}
l(\vecvarthet)&=&%\sum_{i=1}^n\sum_{g=1}^G z_{ig} \log [\pi_g f_t(\vecx_i| \lambda_g, \omega_g, \bm\mu_g, \bm\Sigma_g, \bm\alpha_g)] \notag\\
%&=& 
\sum_{i=1}^n\sum_{g=1}^G z_{ig} \left[ \log (\pi_g) +\sum_{j=1} ^{p} \log [ \phi(\vecx_i | \vecmu_g+y_{ig}\vecalpha_g, y_{ig}\bm\Sigma_g) ] + \log [ h(y_{ig} | \omega_g, \lambda_g ) ]\right],
\label{eq:loglikecluster}
\end{eqnarray}
where $\vecX_i~|~y_{ig} \sim \mathcal{N}(\vecmu_g+y_{ig}\vecalpha_g, y_{ig}\bm\Sigma_g)$ and $Y_{ig} \sim \text{GIG}(\omega_g, 1, \lambda_g)$.
Two steps are iterated until convergence is reached. In the expectation step (E-step), the expected value of the complete-data log-likelihood is computed. Then in the maximization step (M-step), the expected value of the complete-data log-likelihood is maximized with respect to the model parameters. Extensive details of the EM algorithm for generalized hyperbolic mixtures are given by \cite{browne15}.

Following \cite{bohning94} and \cite{lindsay95}, convergence can be determined based on an asymptotic estimate of the log-likelihood at iteration $k+1$, namely $$l_{\infty}^{(k+1)}=l^k+\frac{l^{(k+1)}-l^{(k)}}{1-a^{(k)}},$$ where $$a^{(k)}=\frac{l^{(k+1)}-l^{(k)}}{l^{(k)}-l^{(k-1)}}$$ denotes the Aitken acceleration \citep{aitken26} at iteration $k$.
The algorithm can be considered to have converged when $l_{\infty}^{(k+1)}-l^{(k)}<\epsilon,$ provided that this difference is positive \citep[cf.][]{mcnicholas10a}.

In our applications (Section~\ref{sec:applications}), we assume that the number of components $G$ is unknown. This is not unusual in real model-based clustering applications, where a criterion is often used to determine $G$. The Bayesian information criterion \citep[BIC;][]{schwarz78} is the most popular choice and is given by $$\text{BIC}=2 l(\vecx,\hat\vecvarthet)-r\log n,$$ where $l(\vecx, \hat\vecvarthet)$ is the maximized (observed) log-likelihood, $\hat\vecvarthet$ denotes the maximum likelihood estimate of~$\vecvarthet$, $r$ represents the number of free parameters, and $n$ is the number of observations.
After convergence, component memberships are usually estimated based on the maximum {\it a posteriori} (MAP) classification given by $\text{MAP}\{{\hat z}_{ig}\}=1$ if $\arg \max_h\{{\hat z}_{ih}\}=g$ and $\text{MAP}\{{\hat z}_{ig}\}=0$ otherwise, for $i=1,\ldots,n$.

\subsection{Model-Based Classification and Discriminant Analysis}

Model-based classification, or partial classification \citep[cf.][]{mclachlan82}, is a semi-supervised analogue of model-based clustering that has historically received much less attention within the literature. However, model-based classification has garnered increased attention over the past few years and some authors \citep[e.g.,][]{dean06,mcnicholas10b} have demonstrated that model-based classification can give excellent performance in real applications. Model-based discriminant analysis \citep{hastie96} is a supervised analogue of model-based clustering that has similarly received much less attention until recently  \citep[e.g.,][]{andrews11c,andrews12}. %Model-based classification and discriminant analysis are best explained through the associated likelihoods.
\cite{fraley02a} discuss their own discriminant analysis approach, i.e., MclustDA, as well as the EDDA approach of \cite{bensmail96}.

Consider the classification scenario where there are $n$ observations, $k$ of which have known labelss. Under a model-based classification framework, all $n$ observations are used to estimate the group memberships for the $n-k$ unlabelled observations. Following \cite{mcnicholas10b}, without loss of generality, we arrange the data so that the first $k$ observations are labelled. Accordingly, the likelihood can be written
\begin{equation}
\mathcal{L}(\vecvarthet~|~\vecx)=\prod_{i=1}^{k} \prod_{g=1}^{G}\left[\pi_g f_{\text{h}}(\vecx_i~|~ \lambda_g, \omega_g, \bm\mu_g, \bm\Sigma_g, \bm\alpha_g) \right]^{z_{ig}}\prod_{j=k+1}^n\sum_{s=1}^G\pi_s f_{\text{h}}(\vecx_j~|~\lambda_s, \omega_s, \bm\mu_s, \bm\Sigma_s, \bm\alpha_s)\,.
\label{eq:likeclass}
\end{equation}
As in the case of model-based clustering, parameter estimation is carried out using the EM algorithm. From \eqref{eq:likecluster} and \eqref{eq:likeclass}, we see that model-based clustering can be viewed as a special case of model-based classification that arises by considering \eqref{eq:likeclass} with $k=0$.

For model-based discriminant analysis, we again have $n$ observations, $k$ of which have known labels. Again, we arrange the data so that the first $k$ observations have known labels; however, instead of using all $n$ observations to estimate the unknown labels, we only use the first $k$ observations (i.e., the labelled observations). 
%%%HERE%%%
First, we form the likelihood
\begin{equation}
\mathcal{L}(\vecvarthet~|~\vecx)=\prod_{i=1}^{k} \prod_{g=1}^{G}\left[\pi_g f_{\text{h}}(\vecx_i~|~\lambda_g, \omega_g, \bm\mu_g, \bm\Sigma_g, \bm\alpha_g) \right]^{z_{ig}}.
\label{eq:likedisc}
\end{equation}
Then, the parameter estimates are computed via the EM algorithm. The resulting \textit{a~posteriori} expected values of the $Z_{ig}$ are used to estimate the membership labels of the remaining $n-k$ observations.
In their discriminant analysis approach, \cite{hastie96} allow multiple Gaussian mixture components per class. \cite{scrucca13} extended the dimension reduction and clustering approach of \cite{scrucca10}, cf.\ Section~\ref{sec:gmmdr}), to a discriminant analysis framework. However, for the discriminant analyses herein, we restrict our HMMDR approach (cf.\ Section~\ref{sec:methodology}) to one component per known class. In part, this is done because of the large number of parameters to be estimated and the relatively small number of observations in the real data sets we consider (cf.\ Section~\ref{sec:real}). However, it is also done because we believe that the flexibility inherent in generalized hyperbolic components makes it far less likely, relative to their Gaussian components, that multiple components would be needed to model a single class. This latter point will be investigated as part of future work (cf.\ Section~\ref{sec:conclusion}).

\subsection{Dimension Reduction and Model-Based Clustering (GMMDR)}\label{sec:gmmdr}

 \cite{scrucca10} proposed a method of dimension reduction for model-based clustering within the Gaussian mixture framework, called GMMDR. Given a $G$-component Gaussian mixture model (GMM), i.e., 
\beq 
f(\vecx~|~\vecvarthet)=\sum_{g=1}^G \pi_g f_g(\vecx~|~\vecmu_g, \matsig_g)=\sum_{g=1}^G \pi_g \left[\frac{\exp\left\{-\frac{1}{2}(\vecx - \vecmu_g)^{\top}\matsig_g^{-1}(\vecx - \vecmu_g)\right\}}{(2\pi)^{\frac{p}{2}}|\bm\Sigma_g|^{\frac{1}{2}}}\right], 
\eeq
the procedure finds the smallest subspace that captures the clustering information contained within the data. The core of the method is to identify those directions where the cluster means $\vecmu_g$ and the cluster covariances $\matsig_g$ vary as much as possible, provided that each direction is $\bm\Sigma$-orthogonal to the others. 

Finding these directions is achieved through the generalized eigen-decomposition of the kernel matrix $\bm M$, defined by \cite{scrucca10} as $\bm M \bm v_i=l_i\bm\Sigma\bm v_i$, where $l_1\geq l_2\geq\cdots\geq l_d > 0$ and $\bm v_i^{\top}\bm\Sigma \bm v_j= 1$ if  $i=j$ and  $\bm v_i^{\top}\bm\Sigma \bm v_j= 0$ otherwise. Note that there are $d\leq p$ directions that span the subspace. This kernel contains the variations in cluster means $$\bm{M}_{\mathrm{I}} = \sum_{g=1}^G \pi_g (\vecmu_g-\bm\mu)(\bm\mu_g-\bm\mu)^{\top}$$ and variations in cluster covariances $$\bm M_{\mathrm{II}} = \sum _{g=1}^G \pi_g(\matsig_g-\bar{\matsig})\matsig^{-1}(\matsig_g-\bar{\matsig})^{\top},$$ such that $\bm M=\bm M_{\mathrm{I}}\matsig^{-1}\bm M_{\mathrm{I}} +\bm M_{\mathrm{II}}$.

Here, $\bm\mu=\sum_{g=1}^G\pi_g\vecmu_g$ is the global mean, $\bm\Sigma=({1}/{n})\sum_{i=1}^n(\vecx_i-\bm\mu)(\vecx_i-\bm\mu)^{\top}$ is the covariance matrix, and $\bar{\matsig}=\sum_{g=1}^G\pi_g\bm{\Sigma}_g$ is the pooled within-cluster covariance matrix. Parameter estimation is carried out using the {\tt mclust} package \citep[cf.][]{fraley99} for {\sf R}. The {\tt mclust} software fits a family of ten Gaussian mixture models, which is a subset of the 14 Gaussian parsimonious clustering models (GPCMs) introduced by \cite{celeux95}. The GPCM family arises from the imposition of various constraints on eigen-decomposed component covariance matrices \citep[cf.][]{banfield93,celeux95,fraley02a}.

\subsection{Non-Gaussian Extensions to GMMDR}

The work of \cite{scrucca10} has already been extended to two non-Gaussian mixture settings. In the context of model-based clustering,  \cite{morris13a} proposed a $t$-distribution analogue of GMMDR, called $t$MMDR. This approach uses the $t$EIGEN family of models \citep{andrews12}, which is a $t$-analogue of the GPCM family of models. The most general, i.e., unconstrained, member of the $t$EIGEN family is a mixture model with component density 
\begin{equation}\label{eqn:tdist}
f_t(\vecx~|~\bm\mu_g, \matsig_g, \nu_g)= \displaystyle \frac{\Gamma(\frac{\nu_g+p}{2}) | \bm\Sigma_g |^{-\frac{1}{2}}}{(\pi\nu_g)^{\frac{p}{2}}\Gamma(\frac{\nu_g}{2})(1+\frac{\delta(\vecx, \bm\mu_g|\bm\Sigma_g)}{\nu_g})^{\frac{\nu_g+p}{2}}} \,,
\end{equation}
where $\bm\mu_g$ is the mean, $\bm\Sigma_g$ is a scale matrix, $\nu_g$ is the number of degrees of freedom, and $\delta(\vecx, \vecmu_g | \matsig_g)$ is defined as before. Of course, a mixture model with component density \eqref{eqn:tdist} is just a mixture of multivariate $t$-distributions, which has been applied for  clustering for some time \citep{mclachlan98,peel00}.

\cite{morris13b} developed an analogue of GMMDR for shifted asymmetric Laplace (SAL) mixtures \citep{franczak12}, named SALMMDR. For SAL mixtures, the component density is   
\begin{equation*}
f_s(\vecx|\vecalpha_g, \vecmu_g, \matsig_g)=\frac{2\ \text{exp}\{(\vecx-\vecmu_g)^{\top} \matsig_g^{-1}\bm\alpha_g\}}{(2\pi)^{p/2}| \matsig_g|^{1/2}} \left( \frac{\delta(\vecx, \vecmu_g | \matsig_g)}{2+\vecalpha_g^{\top} \matsig_g^{-1} \vecalpha_g}\right)^{\nu/2} K_{\nu}(u)\,,
\end{equation*}
with mean $\bm\mu_g$, scale matrix $\bm\Sigma_g$, and skewness $\bm\alpha_g$. Here,  $u=\sqrt{(2+\vecalpha_g^{\top} \matsig_g^{-1} \vecalpha_g)\delta(\vecx, \vecmu_g | \matsig_g)}$, $K_{\nu}$ is the modified Bessel function of the third kind with index $\nu=(2-p)/2$, and $\delta(\vecx, \vecmu_g | \matsig_g)$ is as defined previously. The use of SAL mixtures is effective for clustering data with asymmetric components, and they can perform better than Gaussian mixtures in these cases \citep[cf.][]{franczak12}.

The approach introduced herein (Section~\ref{sec:methodology}) aims to combine the robustness offered by the $t$MMDR approach with the elegance and asymmetry afforded by SALMMDR.

\section{Methodology}\label{sec:methodology}

The dimension reduction approach of \cite{scrucca10} is extended through development of a generalized hyperbolic analogue. We will also develop methods for model-based classification and discriminant analysis for GMMDR and all of the non-Gaussian analogues considered herein. Recently, \cite{scrucca13} also extended GMMDR to model-based discriminant analysis.

Given a generalized hyperbolic mixture (\ref{eq:hdensity}), we wish to find a subspace $\mathcal S(\bm\beta)$ where the cluster means and cluster covariances vary the most. Although $\bm\mu_g$ is a mean and $\bm\Sigma_g$ is a covariance matrix in (\ref{eq:hdensity}), note that they are not the mean and covariance matrix of the random variable $\vecX$ with the density in (\ref{eq:hdensity}), except for the special case where $\vecalpha_g = \bm 0$. The mean of $\vecX$ in \eqref{eq:hdensity} is $\tilde{\vecmu}_g \colonequals \bm\mu_g + \bm\alpha_g$, and the covariance of $\vecX$ is $\tilde{\bm\Sigma}_g \colonequals \bm\Sigma_g+\bm\alpha_g\bm\alpha_g^{\top}$. Thus, we define the kernel matrix $\bm M_{\text{HMM}}$ for generalized hyperbolic mixtures to be
\begin{eqnarray}
\bm M_{\text{HMM}}&=&\sum_{g=1}^G \pi_g (\tilde{\vecmu}_g-\bm\mu)(\tilde{\bm\mu}_g-\bm\mu)^{\top} \matsig^{-1} \sum_{g=1}^G \pi_g (\tilde{\vecmu}_g-\bm\mu)(\tilde{\bm\mu}_g-\bm\mu)^{\top}\notag + \sum _{g=1}^G \pi_g(\tilde{\matsig}_g-\bar{\matsig})\matsig^{-1}(\tilde{\matsig}_g-\bar{\matsig})^{\top},
\label{eq:hkernel}
\end{eqnarray}
where $\bm\Sigma=({1}/{n})\sum_{i=1}^n(\vecx_i-\bm\mu)(\vecx_i-\bm\mu)^{\top}$ denotes the overall covariance matrix and $\bar{\matsig}=\sum_{g=1}^G\pi_g\tilde{\bm{\Sigma}}_g$ is the pooled within-cluster covariance matrix.

{\Prop The directions where the cluster means $\tilde{\vecmu}_g$ and the cluster covariances $\tilde{\matsig}_g$ vary the most are obtained from the eigen-decomposition 
\begin{equation}
\bm M_{\text{HMM}} \bm v_i=l_i\bm\Sigma\bm v_i, 
\label{eq:hkerneleigen}
\end{equation}
where $l_1\geq l_2\geq\cdots\geq l_d > 0$ and $\bm v_i^{\top}\bm\Sigma \bm v_j= 1$ if  $i=j$ and  $\bm v_i^{\top}\bm\Sigma \bm v_j= 0$ otherwise.

The eigenvectors $[\bm v_1, \ldots, \bm v_d]\equiv\bm \beta$, with $d\leq p$, form the basis of the dimension reduction subspace $\mathcal S(\bm\beta)$. These eigenvectors are defined as the HMMDR directions.}

{\Prop  Let $\mathcal S(\bm\beta)$ be the subspace spanned by the HMMDR directions obtained from the eigen-decomposition of $\bm M_{\text{HMM}}$ (\ref{eq:hkerneleigen}).
\begin{enumerate}[i]
\item The projections of the parameters onto $\mathcal S(\bm\beta)$ are given by $\bm \beta^{\top} \tilde{\bm\mu}_g$ and $\bm \beta^{\top}\tilde{\bm\Sigma}_g\bm\beta$, respectively.
\item The projections of the $n\times p$ data matrix $\vecx$ onto the subspace $\mathcal S(\bm\beta)$ are computed from $\vecx\bm \beta$. These projections are defined as the HMMDR variables.
\end{enumerate}
}

For an $n\times p$ data matrix $\vecx$, the kernel $\bm M_{\text{HMM}}$ (\ref{eq:hkernel}) is obtained using the estimates from the fit of an HMM on $\vecx$, via an EM algorithm. Then, the HMMDR directions are calculated from the generalized eigen-decomposition of $\bm M_{\text{HMM}}$ (\ref{eq:hkernel}) with respect to the overall covariance matrix $\bm\Sigma$. The HMMDR directions are ordered based on eigenvalues, which means that directions associated with eigenvalues close to zero can be disregarded in practical applications because clusters will superimpose greatly along these directions. 

Similar to GMMDR, the estimation of the HMMDR variables can be interpreted as feature selection, where the members are reduced through a set of linear combinations of the original variables. It is possible that this set of features contains estimated HMMDR variables that do not offer any clustering information but require parameter estimation. \cite{scrucca10} uses the selection method of \cite{raftery06} to prune the subset of GMMDR features. We follow this approach to select the most appropriate HMMDR variables. Two subsets of features, $s$ and $s'=\{ s\setminus  i\}\subset s$, for example, can be compared using the BIC difference
\begin{eqnarray}
\text{BIC}_{\text{diff}}(Z_{i\in s})&=&\text{BIC}_{\text{clust}}(Z_{ s})-\text{BIC}_{\text{not clust}}(Z_{s})
=\underbrace{\text{BIC}_{\text{clust}}(Z_{ s})}_{1}-[\underbrace{\text{BIC}_{\text{clust}}(Z_{ s'})}_{2}+\underbrace{\text{BIC}_{\text{reg}}(Z_i | Z_{ s'})}_{3}],
\label{eq:biccrit}
\end{eqnarray}
where term $1$ in (\ref{eq:biccrit}) denotes the BIC value for the best clustering model fitted using features in~$s$, term $2$ denotes the BIC value for the best clustering model fitted using features in~$s'$, and term $3$ denotes the BIC value for the regression of the $i$th feature on the remaining features in $s'$.

Because the space of all possible subsets contains $2^d-1$ elements, where $d\leq p$, a full feature search is not usually feasible. To this end, we employ the forward greedy search algorithm of \cite{scrucca10} to find a local optimum in the model space. The procedure is based on the forward-backward search algorithm of \cite{raftery06}; however, a backward step is not necessary here because the HMMDR variables are $\bm\Sigma$-orthogonal. The method can be summarized into three main stages. The initial step selects the first feature that maximizes the BIC difference in (\ref{eq:biccrit}), between the best clustering model and the model that assumes no clustering, i.e., a single component. The following step selects the next feature amongst those not previously included to be the one that maximizes the BIC difference in (\ref{eq:biccrit}). This process is iterated until all of the BIC differences for the inclusion of a variable become negative.

At each stage, the search over the model space is performed with respect to the model parameterization and the number of clusters. This algorithm can be applied to the three frameworks under consideration: model-based clustering, classification, and discriminant analysis, by modifying the likelihood functions (\ref{eq:likecluster}),  (\ref{eq:likeclass}), and (\ref{eq:likedisc}), respectively, via the EM procedure. We can now summarize our new method, which we call HMMDR:
\begin{enumerate}
\item Fit an HMM (\ref{eq:hmm}) to the data using the EM algorithm. 
\item Estimate the HMMDR directions: identify directions where the cluster means and cluster variances vary the most, provided each direction is $\matsig$-orthogonal to the others. This is done through the eigen-decomposition of the kernel matrix $\bm M_{\text{HMM}}$ in (\ref{eq:hkernel}). 
\item Select the HMMDR variables: compute the set of features by projecting the data onto the estimated subspace and use the greedy search algorithm to discard the ones that provide no clustering information.
\item Fit an HMM (\ref{eq:hmm}) on the selected HMMDR variables and return to step 2.
\item Repeat steps 2--4 until none of the features can be discarded.
\end{enumerate}

Note that in the analyses herein (Section~\ref{sec:applications}) the {\tt hclust()} from the {\tt mclust} package \citep{fraley12b} for {\sf R} is used for initialization of the EM algorithm in step~1.

%\FloatBarrier
\section{Applications}\label{sec:applications}

\subsection{Performance assessment}

Although the examples herein are treated as genuine clustering and classification analyses, we know the true class labels in all cases. Therefore, we can compare our predicted classifications to the true class labels in each case. To do this, we use the adjusted Rand index \citep[ARI;][]{hubert85}, which is the Rand index \citep{rand71} corrected for chance agreement. The Rand index is based on pairwise agreement and takes a value between $0$ and $1$, where $1$ indicates perfect agreement between two partitions. The correction that leads to the ARI accounts for the fact that random classification is expected to result in some correct agreements; accordingly, the ARI has an expected value of $0$ under random classification and, as with the Rand index, perfect classification corresponds to a value of~$1$. Negative ARI values are possible and indicate classification results worse than would be expected under random classification.

\subsection{Simulated data}
 
\begin{figure}[!ht]
\centering
\includegraphics[width=8cm]{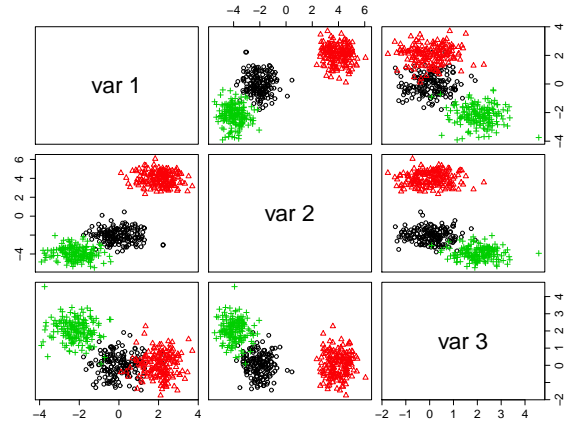}
\vspace{-0.2cm}
\caption{Pairs plot illustrating a generated data set from Scenario~I.}
\label{fig:sim-nonoise}
\end{figure}

\begin{figure}[!ht]
\centering
\includegraphics[width=14cm]{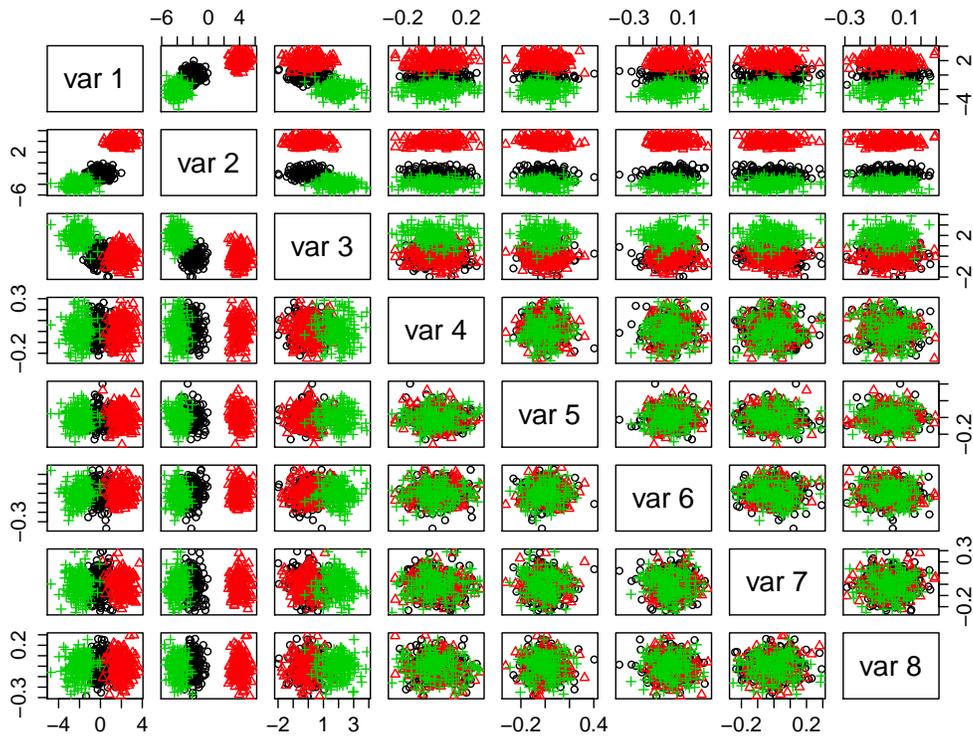}
\vspace{-0.2cm}
\caption{Pairs plot illustrating a generated data set from Scenario~II.}
\label{fig:sim-noise}
\end{figure}

First, we employ a data simulation scheme based on two scenarios to test the HMMDR algorithm and compare it to its Gaussian analogue. In Scenario~I, we generate three variables from a mixture of multivariate Gaussian distributions with mixing proportions $\pi_1=\pi_2=\pi_3=1/3$, means $\bm\mu_1=(0, -2, 0)'$, $\bm\mu_2=(2, 4, 0)'$, $\bm\mu_3=(-2, -4, 2)'$, common covariance matrix $\bm\Sigma=0.5\ident_3$, where $\ident_3$ is the $3\times{3}$ identity matrix, and three different sample sizes $n\in\{100, 500,1000\}$ (e.g., Figure~\ref{fig:sim-nonoise}). In Scenario~II, we modify Scenario~I by adding five noise variables generated from standard normal distributions (e.g., Figure~\ref{fig:sim-noise}).

We run each scenario $300$ times for each possible combination of data dimension and model framework. For model-based classification and discriminant analysis, we assumed that each observation had a $50\%$ probability of being known. This resulted in the number of known observations $k$ being close to $n/2$ but varying slightly from run to run (more details on the selection of known observations appear in the next section). The ARI values for both model-based classification and discriminant analysis was computed based only on the unlabelled observations. The results (Tables~\ref{tab:sim-nonoise} and~\ref{tab:sim-noise}) show that HMMDR generally exhibits high ARI values across both scenarios and all sample sizes for model-based clustering, classification, and discriminant analysis. We notice a slight drop in classification performance for Scenario~II with $100$ observations; however, performance on noisy data improves for larger values of~$n$. One or two features are selected throughout Scenario~I, and, as expected, the addition of noise sometimes leads to a slight increase in the number of features in Scenario~II. 
\begin{table}[!ht]
\caption{Summary of results for the HMMDR and GMMDR approaches on the simulated data from Scenario~I, based on $300$ runs.}
\vspace{-0.2cm}
\centering
{\small
\begin{tabular*}{1\textwidth}{@{\extracolsep{\fill}} ll | rrr |  rrr} 
\hline
%\multicolumn{2}{c |}{Scenario I (no noise)}
 & & \multicolumn{3}{c |}{HMMDR} & \multicolumn{3}{c}{GMMDR}  \\
\multicolumn{1}{l}{} &\multicolumn{1}{l |}{} & $n=100$ & $n=500$ & \multicolumn{1}{l |}{$n=1000$} & $n=100$ & $n=500$ & \multicolumn{1}{l}{$n=1000$} \\
\hline
Clustering  &Avg.\ ARI      &0.9701   &0.9634    &0.9634                    &0.9637  &0.9663    &0.9668\\
            &Std.\ dev.\ ARI     &0.0349   &0.0173    &0.0125                    &0.0347  &0.0166    &0.0111\\
            &Features       &1--2     &1--2      &1                         &1       &1         &1\\
            &Avg.\ no.\ features    &1.137    &1.003     &1                         &1       &1         &1\\
\hline
Classifica- &Avg.\ ARI      &0.9877   &0.9829    &0.9762                    &0.9639  &0.9839    &0.9872\\
tion        &Std.\ dev.\ ARI     &0.0223   &0.0123    &0.0103                    &0.0433  &0.0010    &0.0008\\
            &Features       &1--2     &1--2      &1--2                      &1       &1         &1\\
            &Avg.\ no.\ features    &1.07     &1.033     &1.02                      &1       &1         &1\\
\hline
Discrimin-  &Avg.ARI        &0.9274   &0.9815    &0.9744                    &0.8742  &0.9656    &0.9602\\
ant analysis   &Std.\ dev.\ ARI     &0.1031   &0.0181    &0.0193                    &0.1108  &0.046     &0.038\\
            &Features       &1--2     &1--2      &1--2                      &1--2    &1         &1\\
            &Avg.\ no.\ features    &1.303    &1.027     &1.01                      &1.11    &1         &1\\
\hline
\end{tabular*}
}
\label{tab:sim-nonoise}
\end{table}
\begin{table}[!ht]
\caption{Summary of results for the HMMDR and GMMDR approaches on the simulated data from Scenario~II,  based on $300$ runs.}
\vspace{-0.2cm}
\centering
{\small
\begin{tabular*}{1\textwidth}{@{\extracolsep{\fill}} ll | rrr |  rrr} 
\hline
%\multicolumn{2}{c |}{Scenario II (noise)}
&  & \multicolumn{3}{c |}{HMMDR} & \multicolumn{3}{c}{GMMDR}  \\
\multicolumn{1}{l}{} &\multicolumn{1}{l |}{} & $n=100$ & $n=500$ & \multicolumn{1}{l |}{$n=1000$} & $n=100$ & $n=500$ & \multicolumn{1}{l}{$n=1000$} \\
\hline
Clustering  &Avg.\ ARI      &0.9239   &0.9810    &0.9827                    &0.9632  &0.9655    &0.9656\\
            &Std.\ dev.\ ARI     &0.0692   &0.0120    &0.0072                    &0.0396  &0.0149    &0.0105\\
            &Features       &1--4     &1--3      &1--3                      &1,4     &1         &1\\
            &Avg.\ no.\ features    &2.609    &2.287     &2.277                     &1.021   &1         &1\\
\hline
Classifica- &Avg.\ ARI      &0.9507   &0.9296    &0.9616                    &0.9644  &0.9829    &0.9830\\
tion        &Std.\ dev.\ ARI     &0.0533   &0.0129    &0.0078                    &0.0472  &0.0192    &0.0072\\
            &Features       &1--4     &1--3      &1--3                      &1,4     &1         &1\\
            &Avg.\ no.\ features    &1.07     &1.65      &1.74                      &1.01    &3         &3\\
\hline
Discrimin-  &Avg.ARI        &0.7299   &0.9662    &0.9877                    &0.8336  &0.9634    &0.9609\\
ant analysis   &Std.\ dev.\ ARI     &0.1280   &0.0247    &0.0072                    &0.0978  &0.0354    &0.0316\\
            &Features       &1--3     &1--3      &1--3                      &1--3    &1--3      &1--2\\
            &Avg.\ no.\ features    &1.557    &1.97      &1.797                     &2.425   &1.66      &1.137\\
\hline
\end{tabular*}
}
\label{tab:sim-noise}
\end{table}

The results in Tables~\ref{tab:sim-nonoise} and~\ref{tab:sim-noise} demonstrate that the HMMDR approach gives excellent performance --- for model-based clustering, classification, and discriminant analysis --- for the data from Scenarios~I and~II. However, it is also helpful to have a sense of how long the algorithms take to run. Consider the cases with $n=1000$. For clustering, one run of the HMMDR approach takes as average of 18.5~s in Scenario~I and 37.7~s in Scenario~II. For classification, the equivalent times are 15.4~s and 26.1~s, and for discriminant analysis, the times are 15.0~s and 31.3~s. While it is true that our approach would be considered slow when compared some competitors, these times show that the algorithm does not take a prohibitively long time to run.

\begin{figure}[!ht]
\centering
\includegraphics[width=10cm]{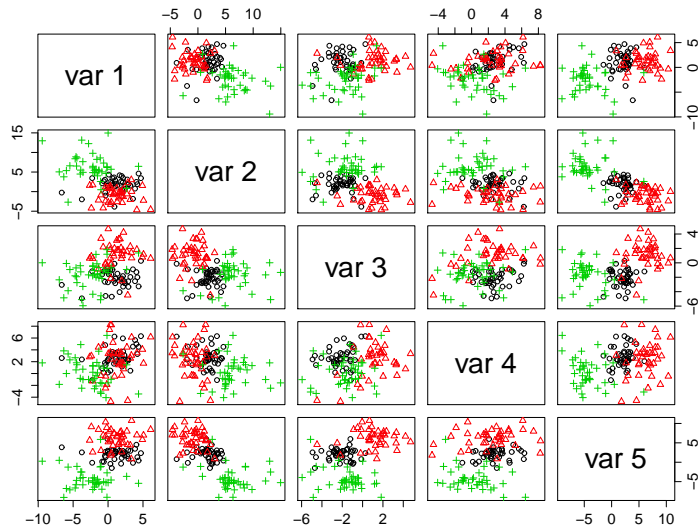}
\caption{Pairs plot illustrating a generated data set from Scenario~III.}
\label{fig:sim-hd}
\end{figure}

In Scenario III, we compare the performance of HMMDR against the existing GMMDR approach for higher dimensional data, including the case where $n_g<p$. We generated three-component data sets with $n_g=40$ observations per component from generalized hyperbolic distributions with random covariance matrices \citep[produced using the {\sf R} package {\tt clusterGeneration};][]{qiu06} and $\vecalpha = -\mathbf{1}$. The means were drawn from a multivariate standard normal distribution and multiplied by a small integer. A typical data set from Scenario~III is illustrated in Figure~\ref{fig:sim-hd}, and clearly this is a difficult clustering problem.
%
%% PAUL HERE
Runs were performed with $p \in \{20,30,40,50\}$, most completed successfully (i.e., converged) but some did not. %some resulted in error due to lack of convergence, combining of groups, or other numerical errors. 
Looking at the results (Table \ref{tab:simhd}), we see that HMMDR tends to produce more successful runs than GMMDR when $p \leq n_g$. We note numerical difficulties for $p > n_g$, where the HMMDR scale matrices will generally be numerically singular; this problem is mitigated in GMMDR because the parsimonious covariance structures from {\tt mclust} are used. When successful, HMMDR performs well in higher dimensions ($p=40,50$) but requires more features than GMMDR. 
\begin{table}[!ht]
\caption{Summary of clustering results for the HMMDR and GMMDR approaches on the simulated data from Scenario~III.}
\vspace{-0.2cm}
\centering{
\begin{tabular}{l c | cc cc cc } 
\hline
         &$p$ & Runs & Succ. & Avg. ARI & Med. ARI & Avg. Feat. & Avg. Comp. \\
\hline
HMMDR          & 20  & 30   & 30    & 0.43     & 0.41     & 9.8        & 2.93       \\
               & 30  & 30   & 29    & 0.72     & 0.88     & 19         & 3          \\
               & 40  & 50   & 44    & 0.94     & 0.96     & 25.1       & 3          \\
               & 50  & 60   & 31    & 0.97     & 1        & 29.5       & 3          \\
\hline
GMMDR          & 20  & 30   & 26    & 0.88     & 0.91     & 4.7        & 3.2        \\ 
               & 30  & 30   & 26    & 0.96     & 1        & 3.2        & 3.1        \\
               & 40  & 50   & 43    & 0.98     & 1        & 3.8        & 3          \\
               & 50  & 60   & 50    & 0.99     & 1        & 3.7        & 3          \\
\hline
\end{tabular}}
\label{tab:simhd}
\end{table}

We note that Scenario~III is not a fair comparison because GMMDR has the massive parsimony advantage of drawing on the {\tt mclust} covariance structures. Repeating the simulations under this scenario with GMMDR restricted to the ``VVV" model, i.e., without any reduction in the number of free covariance parameters, would tell a different story. Of course, implementing the GPCM covariance structures on the scale matrices within the HMMDR approach would also lead to different results. Furthermore, we anticipate that the performance of HMMDR would also improve with the addition of other methods to avoid numerical singularities in the estimation of the scale matrices.

\FloatBarrier
\subsection{Real data}\label{sec:real}
To gauge the performance of our algorithm on real data, we compare results with the eight methods outlined below. Except for $k$-means, we choose these particular comparator methods because they provide model-based analyses while implicitly reducing the dimensionality.
\begin{enumerate}
\item Robust principal component analysis \citep[ROBPCA;][]{hubert05} paired with $t$-mixtures via the $t$EIGEN family: principal components analysis resistant to outliers, with robust loadings computed by using projection-pursuit techniques and the minimum covariance determinant method. We use the {\sf R} package {\tt rrcov} \citep{todorov09} for the ROBPCA computations as well as the {\tt teigen} package  \citep{teigen}.
\item The family of parsimonious Gaussian mixture models \citep{mcnicholas08}, which contains the mixture of factor analyzers model and variants thereof. The {\sf R} package {\tt pgmm} \citep{pgmm} is used. 
\item Mixtures of common factor analyzers \citep{baek10} using the the {\sf R} package {\tt mcfa} \citep{mcfa}.
\item FisherEM \citep{bouveyron12}: a subspace clustering method based on Gaussian mixtures, where an EM-like algorithm estimates both the discriminative subspace and the parameters of the model. The {\sf R} package {\tt FisherEM} \citep{bouveyron12} is employed.
\item The {\sf R} package {\tt clustvarsel} \citep{scrucca13b}.
\item $k$-means clustering using the {\sf R} function {\tt kmeans}.
\item GMMDR: the approach of \cite{scrucca10} based on Gaussian mixtures. While the dimension reduction procedure of GMMDR is available in the {\sf R} package {\tt mclust}, the subset selection procedure is not currently available.
\item $t$MMDR \citep{morris13a}: the $t$-analogue of GMMDR. Fitting of the $t$-mixtures was carried out with the {\sf R} package {\tt teigen}.
\item SALMMDR \citep{morris13b}: the SAL analogue of GMMDR. 
\end{enumerate}

For the analyses in this section, we fit HMMDR and comparator methods to the scaled version of each data set. Where appropriate, we initialize the algorithms with the Gaussian hierarchical agglomerative procedure from {\tt mclust}. In the case of HMMDR and its analogues, we allow the number of components to vary between $G=1$ and $G=6$. Note that we use the term `analogue' somewhat loosely here, because we do not consider decomposed covariance structures for either generalized hyperbolic mixtures or shifted asymmetric Laplace mixtures (Table~\ref{tab:covmethods}).
\begin{table}[!ht]
\caption{Details about the four MMDR methods.}
\vspace{-0.2cm}
\centering{
\begin{tabular*}{0.69\textwidth}{l|rrr}
\hline
          Method & Covariance & Eigen-decomposed $\matsig_g$ & Model family\\ \hline
HMMDR &$\bm\Sigma_g+\bm\alpha_g\bm\alpha_g^{\top}$ &No &--\\
SALMMDR &$\bm\Sigma_g+\bm\alpha_g\bm\alpha_g^{\top}$&No &--\\
$t$MMDR &$\frac{\nu_g }{\nu_g - 2} \matsig_g$, $\nu_g > 2$ &Yes & $t$EIGEN \\
GMMDR &$\bm\Sigma_g$ &Yes & MCLUST \\
\hline
\end{tabular*}}
\label{tab:covmethods}
\end{table}

In the context of model-based classification and discriminant analysis, we use the approach of \cite{mcnicholas10b} to simulate a situation in which some of the group memberships are unknown. For each observation $\vecx_i$, a random number is generated from a uniform distribution on $[0,1]$. If the random number is less than $0.5$, then $\vecx_i$ is taken as known; otherwise, $\vecx_i$ is taken as unknown. To make sure that all the classes are represented, we repeated this procedure for each group until at least one known observation was produced before moving onto the next group. Of course, it follows from this procedure that the number of unknown observations varied from run to run.

We utilize the functionality of {\tt teigen} to fit both Gaussian mixtures and $t$-mixtures for model-based classification \citep[cf.][]{andrews12}. Similarly, {\tt teigen} and {\tt mclust} were employed for model-based discriminant analysis. For each data set, the procedures were run $25$ times, using hierarchical agglomerative starting values of the unknown $\hat{z}_{ig}$. 
Note that we choose each real data set on the basis that it has previously been used to illustrate the performance of some of the comparator methods. We consider that this approach facilitates a very fair comparison. In all cases, we illustrate model-based clustering, classification, and discriminant analysis. For model-based classification and discriminant analysis, the ARI is computed based only on unlabelled observations. We perform random subset cross-validation, training on $25$ different subsets consisting of roughly half the number of observations. This is more challenging than other well known procedures such as $10$-fold cross-validation.   

\subsubsection{Swiss Bank Notes}

%First, we consider two data sets for which HMMDR produced perfect classification results. 
\cite{flury88} present six measurements (length, diagonal, left, right, top, and bottom) taken from genuine and counterfeit Swiss bank notes. These data are available through the {\sf R} package {\ttfamily gclus} \citep{gclus}.
In terms of model-based clustering, HMMDR and its comparators were fitted to these data and the resulting MAP classifications show very high ARI values for most methods (Table~\ref{tab:bankmethods}), with HMMDR and $k$-means being the only methods to cluster the bank notes data perfectly. For model-based classification, HMMDR, SALMMDR, and $t$MMDR produce perfect classifications of the unknown observations (Table~\ref{tab:bankclassda}). However, only HMMDR provides perfect model-based discriminant analysis results on the bank notes. We note that HMMDR selected the minimum number of features in all three scenarios, i.e., one feature.
\begin{table}[!ht]
\centering
\caption{Summary of model-based clustering results for the Swiss bank notes and female voles data sets.}
\begin{tabular*}{1\textwidth}{@{\extracolsep{\fill}}lccccccc}
        \hline
         &\multicolumn{3}{c}{Bank Notes}&&\multicolumn{3}{c}{Female Voles}\\
         \cline{2-4}\cline{6-8}
          & ARI & Features & Components && ARI & Features & Components\\ \hline
HMMDR &1 &1 &2&&1 &1 &2\\
SALMMDR &0.98 &3 &2&&0.95 &2 &2\\
$t$MMDR &0.98&2&2&&0.91 &1 &2\\
GMMDR &0.98& 2 &2&&0.91 &1 &2\\
ROBPCA &0.98 & 4&2&&0.91 &3 &2\\
{\tt FisherEM} &0.98 &1 &2&&0.66 &1 &2\\
{\tt clustvarsel}&0.85&4&3&&0.91 & 3&2\\
{\tt mcfa}&0.98&2&2&&0.91&2&2\\
{\tt pgmm} &0.82 &2 &4&&0.91 &1 &2\\
{\tt kmeans}&1 &-- &2&&0.74&--&2\\ \hline
\end{tabular*} \label{tab:bankmethods}
\end{table}

It is notable that some of the clustering approaches based on a Gaussian mixture, e.g., {\tt clustvarsel} and {\tt pgmm}, return $G> 2$ components (Table~\ref{tab:bankmethods}). In this context, it is interesting that GMMDR returns a $G=2$ component solution. The fact that HMMDR returns $G=2$ components is less surprising because \cite{tortora15b} fit four different non-Gaussian mixture approaches, including a mixture of generalized hyperbolic distributions, to the {\tt banknote} data and choose a model with $G=2$ components in all four cases. Using another non-Gaussian mixture approach, \cite{franczak15} also find a $G=2$ component solution.

\subsubsection{Female Voles}

\citet{flury97} discuss seven measurements (Table~\ref{tab:volevariables}) of female voles from two species ({\em Microtus californicus} and {\em Microtus ochrogaster}) originally studied by \cite{airoldi84}. The data are available within the {\sf R} package {\tt Flury} \citep{flury10}. Tables~\ref{tab:bankmethods} and \ref{tab:bankclassda} indicate that, out of all of the procedures fitted to the voles data, HMMDR is the only one giving perfect classification results in all three paradigms. 
\begin{table}[!htb]
\caption{Measurements taken for the female vole data.}
\vspace{-0.2cm}
\centering
\begin{tabular*}{1\textwidth}{@{\extracolsep{\fill}} l l }
\hline
Age in days 			&Incisive foramen length\\
Condylo incisive length 	&	Skull height	\\	 
Alveolar length of upper molar tooth row & Interorbital width\\	
Zygomatic width\\ \hline
\end{tabular*}
\label{tab:volevariables}
\end{table}

\begin{table}[!ht]
\centering
\caption{Summary of model-based classification and discriminant analysis results for the bank note data, based on $25$ runs.}
\begin{tabular*}{1\textwidth}{@{\extracolsep{\fill}}lccccccc}
\hline
&\multicolumn{3}{c}{Bank Notes}&&\multicolumn{3}{c}{Female Voles}\\
         \cline{2-4}\cline{6-8}
& ARI & Features & Components & & ARI & Features & Components\\ \hline
HMMDR class. &1 &2 &2&&1 &1 &2\\
SALMMDR class. &1 &2 &2&&1 &3 &2\\
$t$MMDR class. &1 &2 &2&&0.95 &1 &2\\
GMMDR class. &0.96 &2 &2&&0.96 &1 &2\\
\hline
\hline
HMMDR DA &1 &2 &2&&1 &2 &2\\
SALMMDR DA &0.96 &2 &2&&0.91 &3 &2\\
$t$MMDR DA &0.95 &1--2 &2&&0.91 &1 &2\\
GMMDR DA &0.98 &1--4 &2&&0.88 &1--4 &2\\
 \hline
\end{tabular*} \label{tab:bankclassda}
\end{table}

\FloatBarrier
\subsubsection{Italian Wines} 
\cite{forina86} recorded several chemical and physical properties for three types of Italian wines: Barolo, Grignolino, and Barbera. As shown in Table~\ref{tab:winevariables}, thirteen properties for $178$ wines are available from the {\sf R} package {\tt gclus} \citep{hurley04}. 
\begin{table}[!htb]
\caption{Chemical and physical properties available for the wine data in {\tt gclus}.}
\vspace{-0.2cm}
\centering{
\begin{tabular*}{0.99\textwidth}{@{\extracolsep{\fill}} lll}
\hline
Alcohol 		&	Proline			&	OD280/OD315 of diluted wines \\
Malic acid 		&	Ash				&	Alkalinity of ash \\
Hue  &	Total phenols		&	Magnesium \\	
Color intensity	&	Nonflavonoid phenols	&	Proanthocyanins \\
Flavonoids&&\\	\hline
\end{tabular*}}
\label{tab:winevariables}
\end{table}

Within the model-based clustering framework, HMMDR is the best performer ($\text{ARI}=0.97$, Table~\ref{tab:winemethods}), with only two misclassified observations (Table~\ref{tab:winecluster}). The model-based classification scenario (Table~\ref{tab:wineclassda}) reveals that HMMDR, SALMMDR, and $t$MMDR produce perfect classification results. With an ARI of 0.92, HMMDR gives the best performance within the model-based discriminant analysis paradigm. %PAUL%\textbf{In terms of model-based discriminant analysis on the unknown wines, HMMDR drops in performance to an $\text{ARI}=0.92$ (Table~\ref{tab:wineclassda}), while selecting a higher number of features than it did for clustering and classification. We note that both SALMMDR and $t$MMDR result in perfect classifications for the observations taken to be unknown.}
\begin{table}[!ht]
\caption{Model-based clustering, classification, and discriminant analysis results for our HMMDR approach fitted to the wine data. Model-based classification and discriminant analysis results are based on $25$ runs.}
\vspace{-0.2cm}
\centering{
\begin{tabular*}{1\textwidth}{@{\extracolsep{\fill}} l ccc c ccc c ccc}
\hline
 & \multicolumn{3}{c}{Clustering}   && \multicolumn{3}{c}{Classification} && \multicolumn{3}{c}{Disc.\ Anal.}\\
\cline{2-4}\cline{6-8}\cline{10-12} 
&{1}&{2}&{3} & &{1}&{2}&{3} && 1 &2 &3\\ 
%\cline{1-1} \cline{2-4}  \cline{6-8} \cline{10-12}
\hline
Barolo	 & 59&0&0& &875 &0&0	&&625	 &50  &0 \\		
Grignolino& 0&68&2 & &0&925&0    &&0   &825  &0  	\\		
Barbera	 &0&0&48 & &  0& 0&500   &&0	  &0  &550 \\
%\cline{1-1} \cline{2-4}  \cline{6-8} \cline{10-12} 
\hline
ARI; Features &\multicolumn{3}{c}{0.97; 7}&&\multicolumn{3}{c}{1; 5} &&\multicolumn{3}{c}{0.92; 8} \\  \hline 
\end{tabular*}}
\label{tab:winecluster}
\end{table}

\begin{table}[!ht]
\centering
\makebox[0pt][c]{\parbox{1\textwidth}{%
    \begin{minipage}[t]{0.45\hsize}\centering{
\caption{Summary of model-based clustering results for the wine data.}
\vspace{-0.2cm}
        \begin{tabular}{l | ccc}
        \hline
          Method & ARI & Features & Components\\ \hline
HMMDR &0.97 &7 &3\\
SALMMDR &0.92 &10 &3\\
$t$MMDR &0.93 &4 & 3\\
GMMDR &0.85& 5 &3\\
ROBPCA &0.83 &4 &3\\
{\tt FisherEM} &0.91 &2 &3\\
{\tt clustvarsel}&0.78 &5 &3\\
{\tt mcfa}&0.90&3&3\\
{\tt pgmm} &0.79&2 &4 \\
{\tt kmeans}&0.90 &- &3\\ \hline
\end{tabular} \label{tab:winemethods}}
       
    \end{minipage}
    \hfill
    \begin{minipage}[t]{0.48\hsize}\centering{
\caption{Summary of model-based classification and discriminant analysis results for the wine data, based on $25$ runs.}
\vspace{-0.2cm}
        \begin{tabular}{l | ccc}
        \hline
          Method & ARI & Features & Components\\ \hline
HMMDR class. &1 &5 &3\\
SALMMDR class. &1 &10 &3\\
$t$MMDR class. &1 &5 &3\\
GMMDR class. &0.93 &3--8 &3\\
\hline
\hline
HMMDR DA &0.92 &8 &3\\
SALMMDR DA &0.88 &8 &3\\
$t$MMDR DA &0.85 &1--8 &3\\
GMMDR DA &0.85 &2--8 &3\\
 \hline
\end{tabular}\label{tab:wineclassda}}
        
    \end{minipage}%
}}
\end{table}

\begin{figure}[!ht]
\centering
~\hspace{-0.75in}\includegraphics[width=11.5cm]{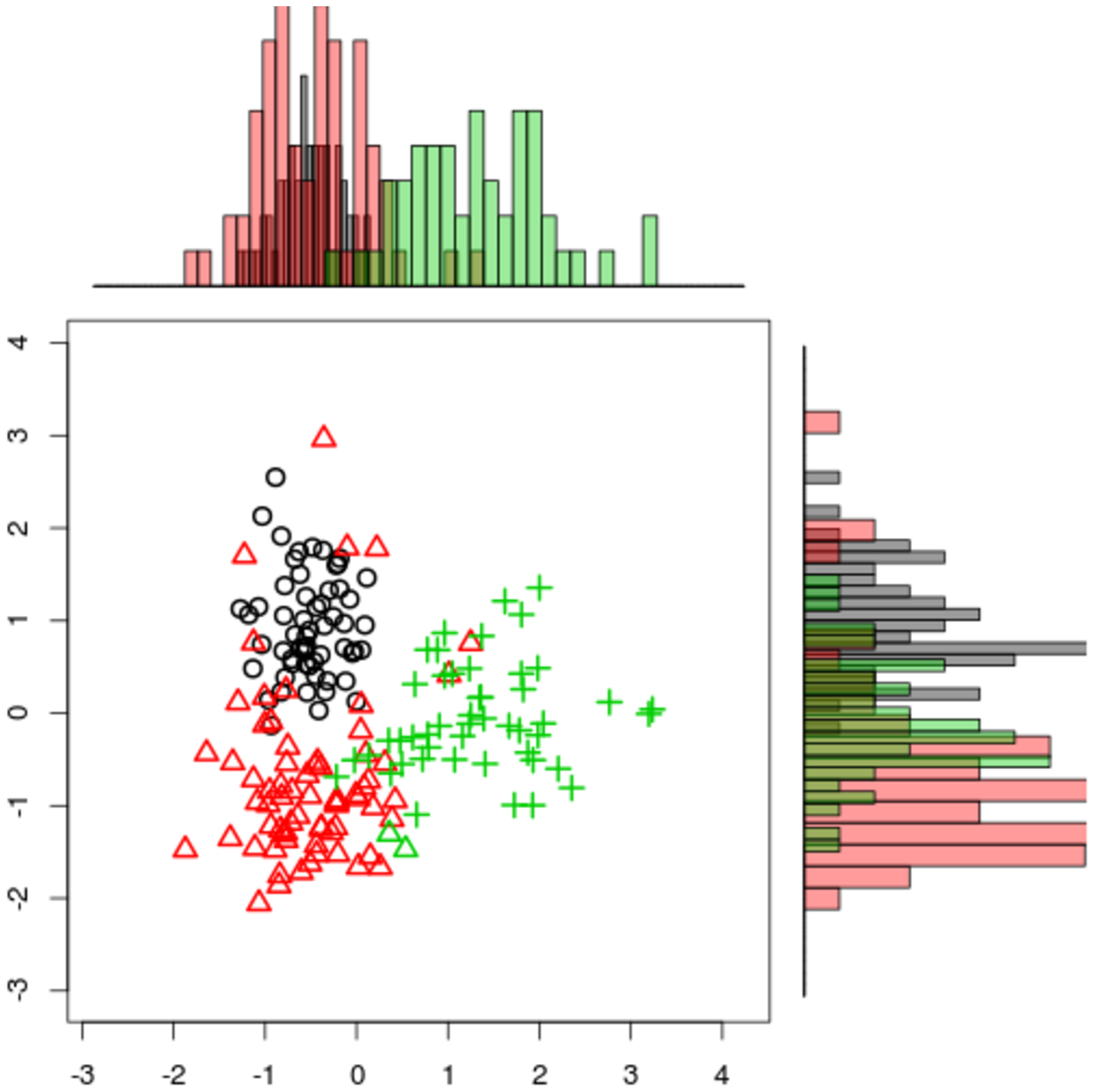}~\hspace{-0.55in}\includegraphics[width=11.5cm]{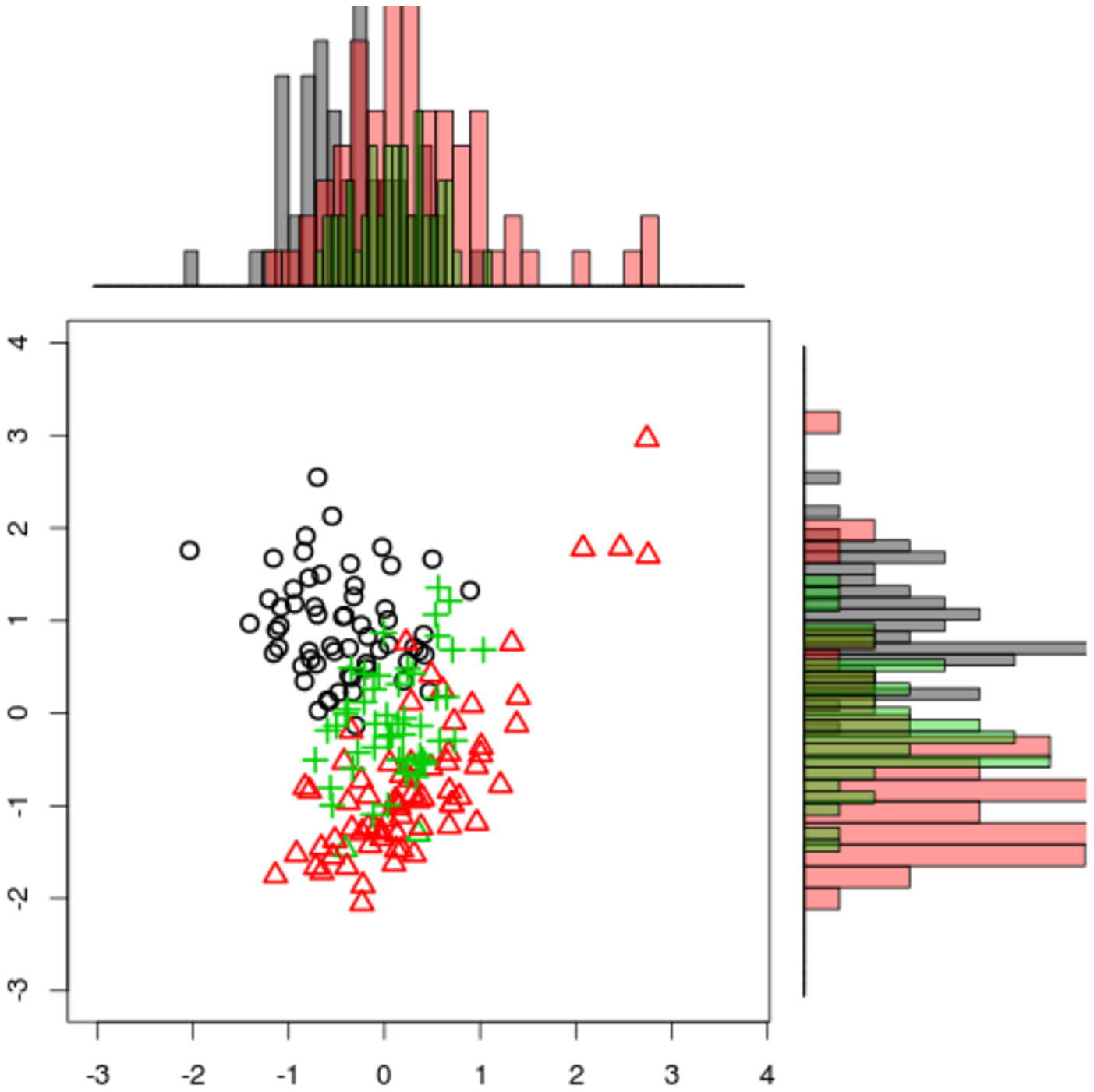}
\caption{Plots of some of the estimated HMMDR directions for the wine data obtained from model-based clustering (directions $2$ vs.\ $4$ on the left-hand side, and directions $4$ vs.\ $5$ on the right-hand side). Symbols indicate true cluster membership and colours indicate the estimated HMMDR cluster allocation. The edge histograms depict the estimated distributions of the observations in each cluster.}
\label{fig:winedirections}
\end{figure}

Figure~\ref{fig:winedirections} illustrates three of the estimated HMMDR directions obtained from our model-based clustering of the wine data (Table~\ref{tab:winecluster}). The edge histograms depict the distribution of the observations from the estimated directions, coloured by estimated cluster allocation. The plot on the left-hand side reveals quite clearly the underlying cluster structure in the data. Although there is some overlap in these two directions, only two wines were misclassified by the HMMDR method and so some of the other dimensions must give additional clarity. 
%We note that, even though some of the observations are somewhat displaced from their respective clusters,  HMMDR was able to cluster them correctly (for example, the few red-coloured observations nestled among the green-coloured ones shown on the left side plot). 
%\enlargethispage{\baselineskip}
%\vspace{-1in}

\FloatBarrier
\subsubsection{Wisconsin Breast Cancer Study}

\cite{mangasarian95} presented a study of breast cancer from Wisconsin, undertaken to establish whether fine needle aspiration of breast tissue samples could classify tumour status. Several attributes are recorded (Table~\ref{tab:wbcavariables}) for $681$ cases of potentially cancerous tumours, of which $238$ were actually malignant. These data are available in the {\sf R} package {\tt faraway} \citep{faraway11}.
\begin{table}[!htb]
\caption{Tissue sample properties of the Wisconsin breast cancer data.}
\vspace{-0.2cm}
\centering{
\begin{tabular*}{1\textwidth}{@{\extracolsep{\fill}}l l l}
\hline
Marginal adhesion 		&	Epithelial cell size			&	Clump thickness \\
Bare nuclei	&	Mitoses				&	Cell shape uniformity \\
Bland chromatin &	Normal nucleoli		&	Cell size uniformity \\	\hline
\end{tabular*}}
\label{tab:wbcavariables}
\end{table}

Within the model-based clustering framework, HMMDR selected three features and gave the best classification performance ($\text{ARI}=0.89$, Table~\ref{tab:wbcacluster}). The $t$MMDR ($\text{ARI}=0.86$) and $k$-means ($\text{ARI}=0.84$) approaches are close behind; however, the rest of the comparators did not produce particularly good results on these data (Table~\ref{tab:wbcamethods}). In the model-based classification and discriminant analysis scenarios, HMMDR gave classification performance similar to clustering and was again the best performer over the 25 runs (Table~\ref{tab:wbcaclassda}). 
\begin{table}[!ht]
\caption{Model-based clustering, classification, and discriminant analysis results for our HMMDR approach on the breast cancer data. Model-based classification and discriminant analysis results are based on $25$ runs.}
\vspace{-0.2cm}
\centering
\begin{tabular*}{1\textwidth}{@{\extracolsep{\fill}} l cc c cc c cc}
\hline
 & \multicolumn{2}{c}{Clustering}   && \multicolumn{2}{c}{Classification} && \multicolumn{2}{c}{Disc.\ Anal.}\\
\cline{2-3}  \cline{5-6} \cline{8-9}
&1&2 & &1&2 && 1&2\\ 
\hline
%\cline{1-1} \cline{2-3}  \cline{5-6} \cline{8-9}
Malignant&236&2 &&2925 &125 &&2575 &200 \\
Benign&17&426  &&100 &4925 &&50  &4675\\ 
\cline{1-1} \cline{2-3}  \cline{5-6} \cline{8-9} 
ARI; Features &\multicolumn{2}{c}{0.89; 3}&&\multicolumn{2}{c}{0.89; 2}&&\multicolumn{2}{c}{0.87; 6}\\ \hline
\end{tabular*}
\label{tab:wbcacluster}
\end{table}
\begin{table}[!ht]
\centering
\makebox[0pt][c]{\parbox{1\textwidth}{%
    \begin{minipage}[b]{0.45\hsize}\centering{
\caption{Summary of model-based clustering results for the breast cancer data.}
\vspace{-0.2cm}
        \begin{tabular}{l | ccc}
        \hline
          Method & ARI & Features & Comp. \\ \hline
HMMDR &0.89 &3 &2\\
SALMMDR &0.80 &5 &2\\
$t$MMDR &0.86 &3 & 2\\
GMMDR &0.58& 5 &2\\
ROBPCA &0.55 &5 &3\\
{\tt FisherEM} &0.79 &1 &2\\
{\tt clustvarsel}&0.78 &2 &2\\
{\tt mcfa}&0.65&3&2\\
{\tt pgmm} &0.42&2 &4 \\
{\tt kmeans}& 0.84&-- &2\\ \hline
\end{tabular}\label{tab:wbcamethods}}
        
    \end{minipage}
    \hfill
    \begin{minipage}[b]{0.48\hsize}\centering{
\caption{Summary of model-based classification and discriminant analysis results for the breast tissue data, based on $25$ runs.}
\vspace{-0.2cm}
        \begin{tabular}{l | ccc}
        \hline
          Method & ARI & Features & Comp. \\ \hline
HMMDR class. &0.89 &2 &2\\
SALMMDR class. &0.85 &4  &2\\
$t$MMDR class. &0.86 &1--2 &2\\
GMMDR class. &0.86 & 1&2\\
\hline
\hline
HMMDR DA &0.87 &6 &2\\
SALMMDR DA &0.84&7 &2\\
$t$MMDR DA & 0.85&2 &2\\
GMMDR DA &0.84 &1--7 &2\\
 \hline
\end{tabular} \label{tab:wbcaclassda}}
        
    \end{minipage}%
}}
\end{table}

\begin{figure}[!ht]
\centering
~\hspace{-0.75in}\includegraphics[width=11.5cm]{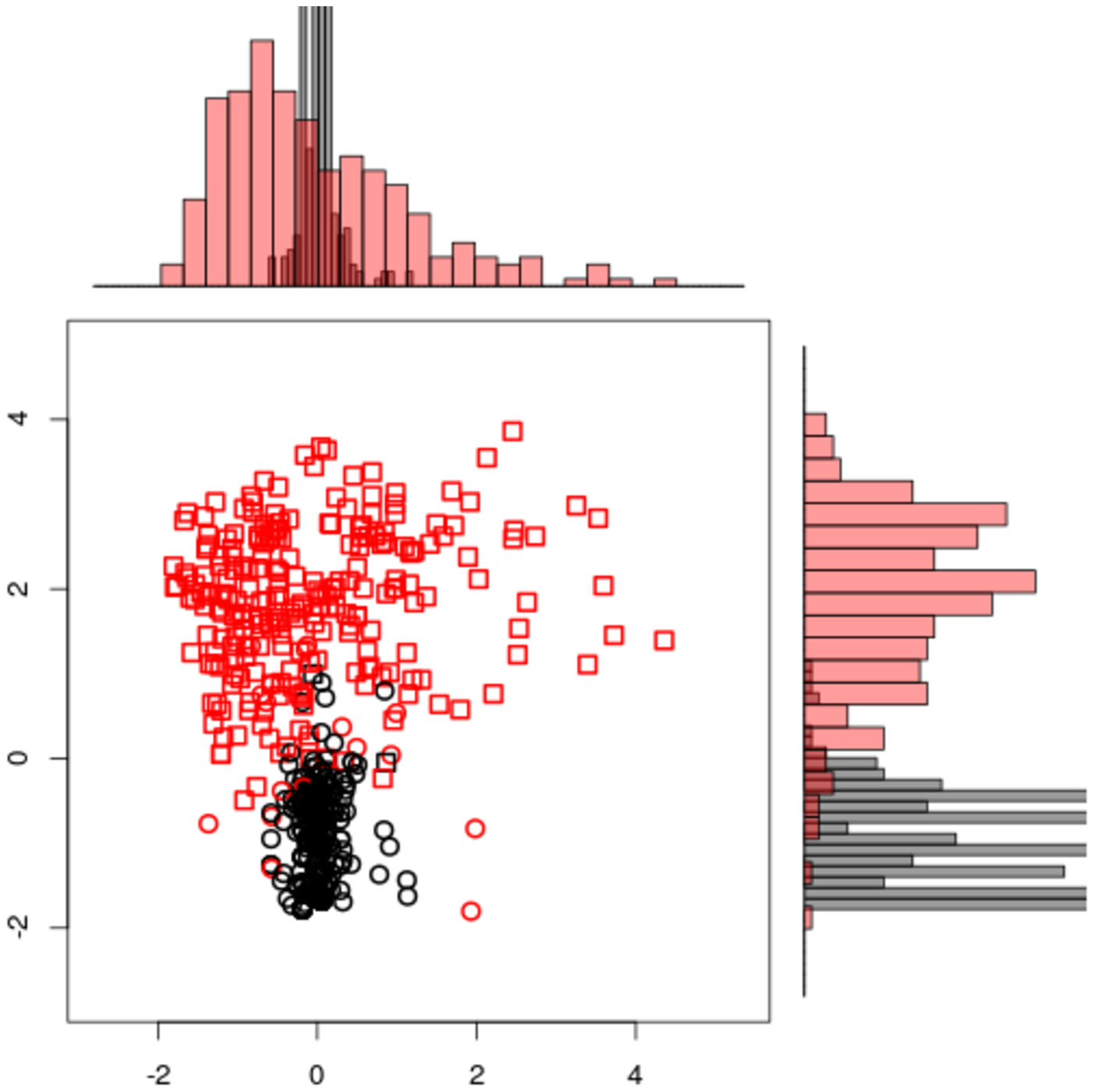}~\hspace{-0.55in}\includegraphics[width=11.5cm]{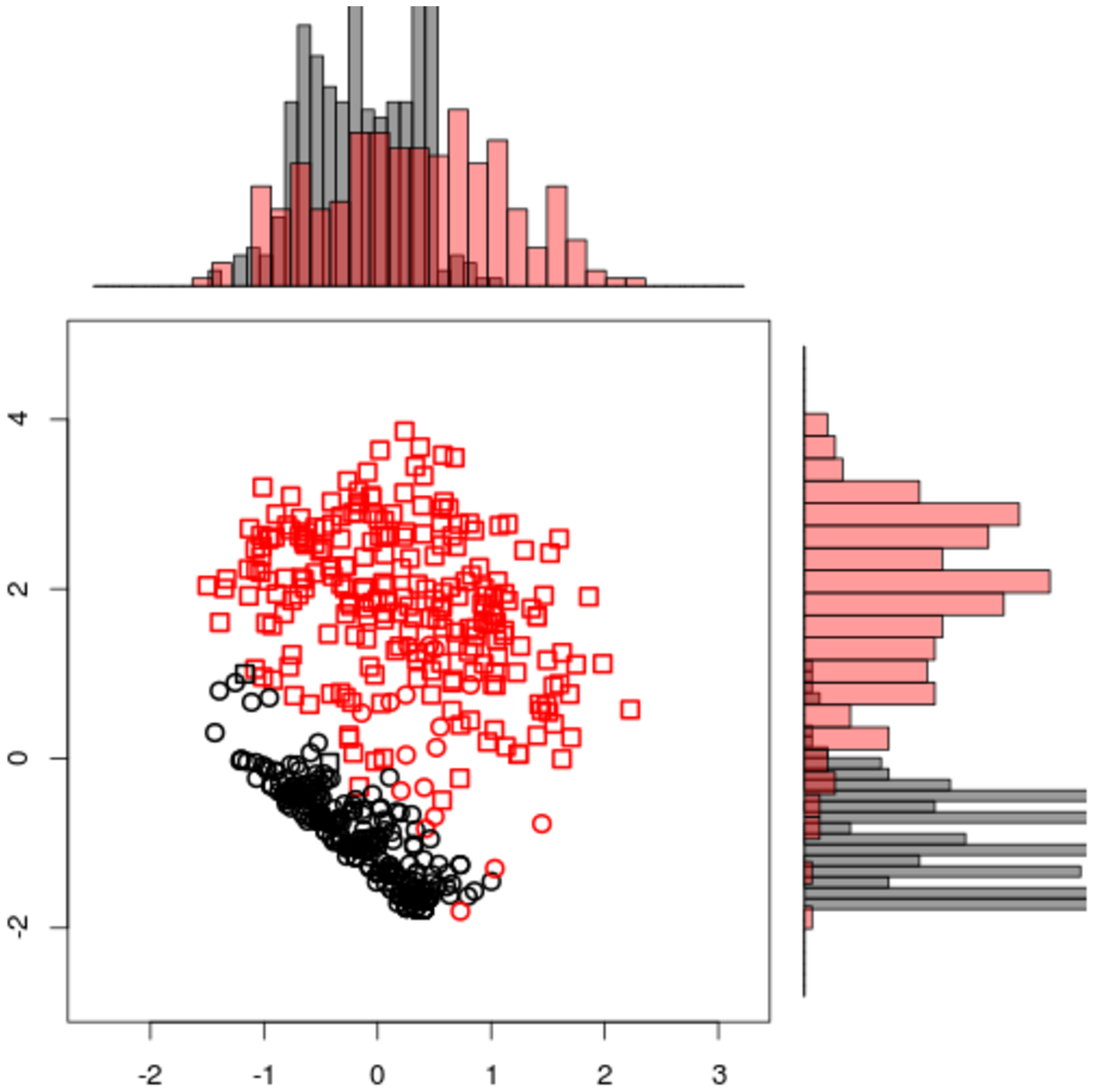}
\caption{Plots of some of the estimated HMMDR directions for the breast cancer data obtained from model-based clustering (direction $1$ vs.\ $3$ on the left-hand side, and direction $2$ vs.\ $3$ on the right-hand side). Symbols indicate true cluster membership and colours indicate the estimated HMMDR cluster allocation. The edge histograms depict the estimated distributions of the observations in each cluster.}
\label{fig:wbcadirections}
\end{figure}

Figure~\ref{fig:wbcadirections} illustrates three of the estimated HMMDR directions obtained from the model-based clustering output shown in  Table~\ref{tab:wbcacluster}. The plots depict quite clearly the inherent cluster structure in the data, and we notice that the malignant breast tissues are quite tightly packed in their cluster. The red histogram on the top edge of the left-hand side of Figure~\ref{fig:wbcadirections} is a nice illustration of a skew distribution along a HMMDR direction.
%\vspace{-1in}

%\FloatBarrier
\subsubsection{Colon Cancer}

\cite{alon99} analyzed gene expression data from microarray experiments of colon tissue, probed by oligonucleotide arrays. A reduced data set containing $62$ tissue samples --- $40$ tumour tissues and $22$ normal tissues --- and $2,000$ genes is available in the {\sf R} package {\tt plsgenomics} \citep{plsgenomics}. 
The challenge here is dealing with the large number of gene expression levels compared with the small number of tissue samples. As is the case with microarray data in general, there are many non-informative genes that can obstruct the clustering of the samples. Thus, gene filtering was carried out prior to further analysis. While \cite{mclachlan02} and \cite{mcnicholas10d} used the EMMIX-GENE procedure \citep{mclachlan02} to reduce the dimensionality of the colon data, we considered a different method.  

Our gene filtering approach is to find differentially expressed genes based on modified $t$-tests, using the {\sf R} Bioconductor package {\tt siggenes} \citep{schwender12}. This package contains the function {\tt sam}, which implements the significance analysis of microarrays (SAM) method proposed by \cite{tusher01}. SAM computes a statistic $d_i$ for each gene $i$, measuring the strength of the relationship between gene expression and the response variable (which is the class variable in our case). It uses repeated permutations of the data to determine if the expression of any gene is significantly related to the response. The cutoff for significance is determined by a tuning parameter $\Delta$, chosen by the user based on the false positive rate. We employed $100$ permutations and chose $\Delta=2.4$, which yielded $23$ genes for analysis (Table~\ref{tab:colonvariables}, \ref{app:colon}).

Even with the dimensionality reduced to $23$ genes, the analysis of the colon data was quite challenging. When HMMDR was fitted to these data within the model-based clustering paradigm, five observations were misclassified (Table~\ref{tab:coloncluster}), corresponding to an ARI of $0.70$. This was the best result, with $t$MMDR ($\text{ARI}=0.64$) being the next best performer (Table~\ref{tab:colonmethods}). For model-based classification and discriminant analysis, HMMDR gave better performance with ARI values of $0.86$ and $0.83$, respectively (Table~\ref{tab:colonclassda}). We note that SALMMDR, $t$MMDR, and GMMDR also gave improved performance within the model-based classification and discriminant analysis paradigms; however, HMMDR was the best approach across all paradigms.
%\enlargethispage{\baselineskip}
\begin{table}[!htb]
\caption{Model-based clustering, classification, and discriminant analysis results for the colon data. Model-based classification and discriminant analysis results are based on $25$ runs.}
\vspace{-0.2cm}
\centering
\begin{tabular*}{1\textwidth}{@{\extracolsep{\fill}} l cc c cc c cc}
\hline
 & \multicolumn{2}{c}{Clustering}   && \multicolumn{2}{c}{Classification} && \multicolumn{2}{c}{Disc.\ Anal.}\\
\cline{2-3}  \cline{5-6} \cline{8-9}
&1&2 & &1&2 && 1&2\\ 
\hline
Normal&22&0 & &225&0 & &220 &15 \\
Tumour&5&35  & &25&425 &&15  &420\\ 
\hline
ARI; Features &\multicolumn{2}{c}{0.70; 3}&&\multicolumn{2}{c}{0.86; 5}&&\multicolumn{2}{c}{0.83; 6}\\ \hline
\end{tabular*}
\label{tab:coloncluster}
\end{table}

\begin{table}[!htb]
\centering
\makebox[0pt][c]{\parbox{1\textwidth}{%
    \begin{minipage}[t]{0.45\hsize}\centering{
\caption{Summary of model-based clustering results for the best models fitted to the colon data.}
\vspace{-0.2cm}
        \begin{tabular}{l | ccc}
        \hline
          Method & ARI & Feat. & Comp. \\ \hline
HMMDR &0.70 &3 &2\\
SALMMDR &0.59 &10 &2\\
$t$MMDR &0.64 &1 &2 \\
GMMDR &0.59&1  &2\\
ROBPCA &0.36 &3 &3\\
{\tt FisherEM} &0.59 &1 &2\\
{\tt clustvarsel}&0.35 &3 &4\\
{\tt mcfa}&0.64&5&2\\
{\tt pgmm} &0.40&3 &2 \\
{\tt kmeans}&0.59 &- &2\\ \hline
\end{tabular}\label{tab:colonmethods}}
        
    \end{minipage}
    \hfill
    \begin{minipage}[t]{0.48\hsize}\centering{
\caption{Summary of model-based classification and discriminant analysis results for the colon data, based on $25$ runs.}
\vspace{-0.2cm}
        \begin{tabular}{l | ccc}
        \hline
          Method & ARI & Feat. & Comp. \\ \hline
HMMDR class. &0.86 &5 &2\\
SALMMDR class. &0.71 &6 &2\\
$t$MMDR class. &0.75 &1 &2\\
GMMDR class. &0.73 &1 &2\\
\hline
\hline
HMMDR DA &0.83 &6 &2\\
SALMMDR DA &0.70 &7 &2\\
$t$MMDR DA &0.74 &1 &2\\
GMMDR DA &0.70 &1 &2\\
 \hline
\end{tabular}\label{tab:colonclassda}}
        
    \end{minipage}%
}}
\end{table}

\begin{figure}[!ht]
\centering
~\hspace{-0.75in}\includegraphics[width=11.5cm]{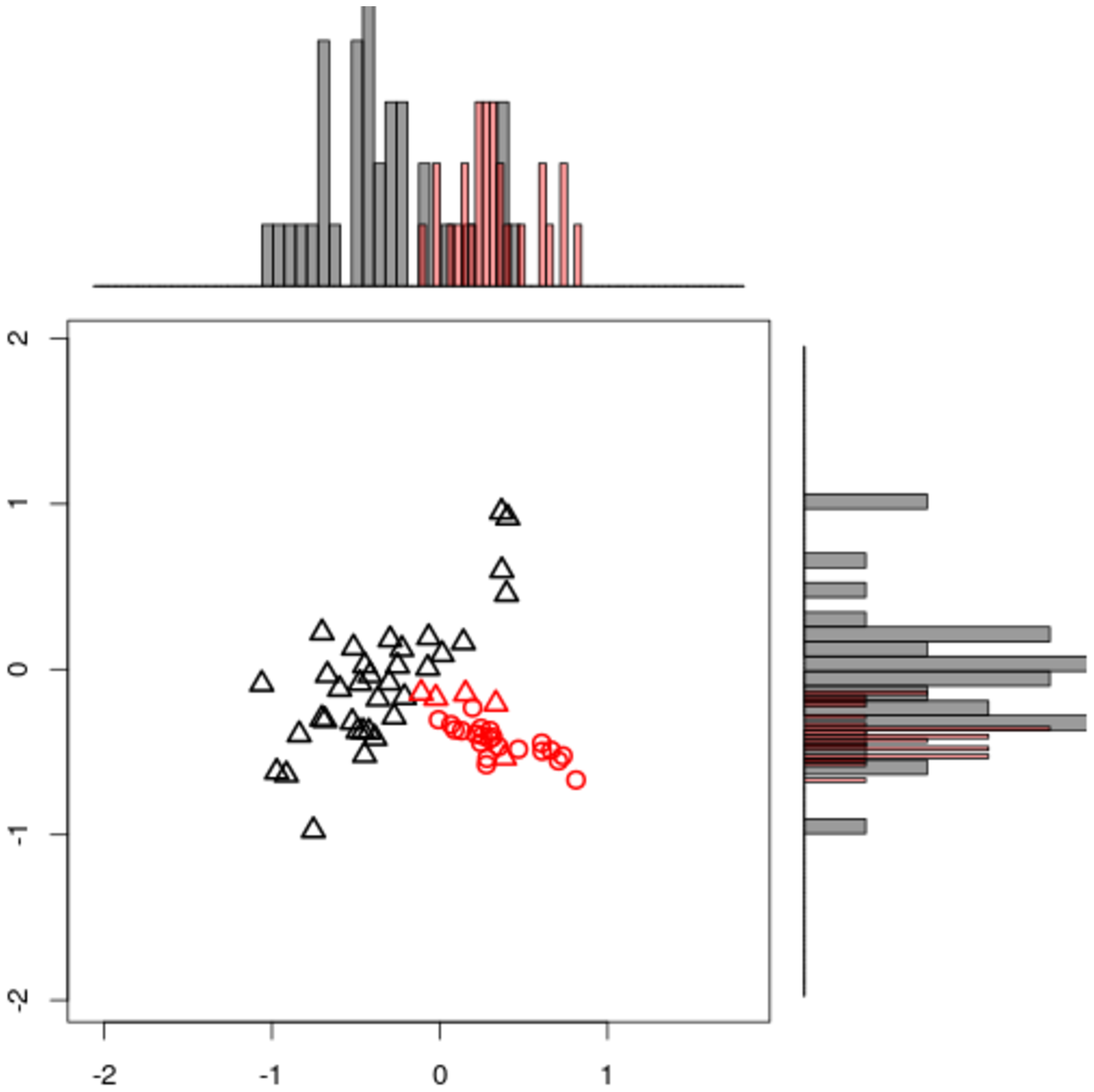}~\hspace{-0.65in}\includegraphics[width=11.5cm]{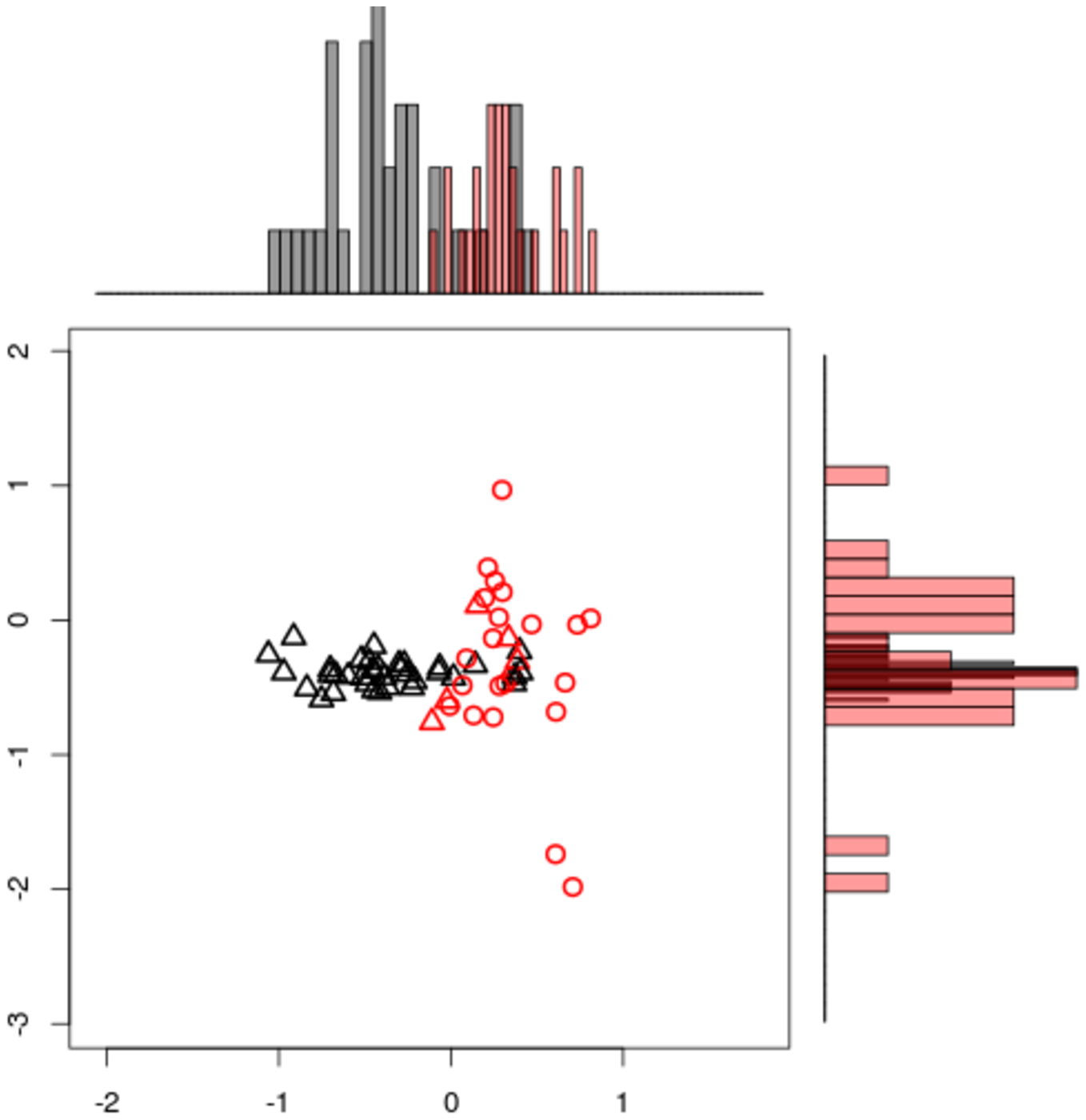}
\caption{Plots of some of the estimated HMMDR directions for the colon data obtained from model-based clustering (direction $2$ vs.\ $3$ on the left-hand side, and direction $1$ vs.\ $2$ on the right-hand side). Symbols indicate true cluster membership and colours indicate the estimated HMMDR cluster allocation. The edge histograms depict the estimated distributions of the observations in each cluster.}
\label{fig:colondirections}
\end{figure}

Figure~\ref{fig:colondirections} illustrates three of the estimated HMMDR directions obtained from the model-based clustering output from Table~\ref{tab:coloncluster}. The plots depict the inherent cluster structure in the colon tissues and, as we would expect, the misclassified tissues are generally close to the cluster boundaries.
%\vspace{-1in}

It is interesting to note that \cite{mcnicholas10d} obtained similar clustering results to HMMDR for the colon data. They considered a subset of $461$ genes and the best model fitted to these data had six latent factors, with five misclassified tissues and an \text{ARI} of $0.70$. Although equal to two significant figures, we point out for completeness that our HMMDR approach has a very slightly higher ARI value (0.699 vs.\ 0.697). However, we must also point out that we used knowledge of tissue type in our gene selection while \cite{mcnicholas10d} did not. \cite{mclachlan02} also analyzed these data; they selected $446$ genes and identified five types of clusterings on this subset. However, these clusterings did not correspond to the tissue type.

\FloatBarrier
\section{Conclusions}\label{sec:conclusion}
This paper introduced an effective dimension reduction technique for model-based clustering, classification, and discriminant analysis using multivariate mixtures of generalized hyperbolic distributions. Our method, known as HMMDR, focused on identifying the smallest subspace of the data that captured the inherent cluster structure. The HMMDR approach was illustrated using simulated and real data, where it performed favourably compared to its existing special cases, i.e., GMMDR, $t$MMDR, and SALMMDR. In clustering applications, HMMDR consistently outperformed several other model-based dimension reduction methods ({\tt ROBPCA}, {\tt pgmm}, {\tt FisherEM},  {\tt clustvarsel}, and {\tt mcfa}). 

One limitation sometimes encountered using our current approach is singularities while fitting full $p \times p$ covariance matrices for each mixture component when there are $n_g<p$ observations for the component. One purely numerical approach we have used to mitigate singularities was to regularize the updated covariance matrices at each iteration in the EM algorithm using a suitably chosen eigenvalue cutoff. Future work will include using analogues of the GPCM models, via eigen-decomposition of the component scale matrices as well as investigating alternative approaches for dealing with singularity problems. The latter might include working on distinct subspaces for fitting each covariance matrix and only injecting into a common subspace when forming $\bm M$.

The real data sets used for our illustrations were selected because they were previously used to illustrate the performance of some of the comparator methods. Therefore, it is encouraging that HMMDR consistently outperformed the comparator methods on real data sets. As part of the development of a companion {\sf R} package, we will study whether %incorporating generalized hyperbolic analogues of the GPCM models into our approach is beneficial. While such an approach is taken within GMMDR and $t$MMDR, we believe that careful study is needed before determining whether adding these analogues is desirable. This also ties in with whether 
multiple components should be available to represent a class in HMMDR discriminant analysis. Finally, the application of our approach within the fractionally-supervised classification framework \citep{vrbik13} will also be a subject of future work.

\section*{Acknowledgements}

The authors are grateful to an associate editor and two anonymous reviewers for their very helpful comments. This work was supported by an Ontario Graduate Scholarship (Morris), an Early Researcher Award from the Government of Ontario (McNicholas), and a Discovery Grant from the Natural Sciences and Engineering Research Council of Canada (NSERC; McNicholas). The computing equipment used was provided through a Research Tools and Instruments Grant from NSERC.  

%\bibliographystyle{apalike}
%\section*{References}
%\bibliographystyle{elsarticle-harv} 
%
%\bibliography{hmmdr}

\newpage

\appendix
\section{Selected Genes for the Colon Cancer Data}\label{app:colon}
\begin{table}[!htb]
\caption{Genes selected with SAM for the colon data.}
\vspace{-0.2cm}
\centering{
\begin{tabular*}{1\textwidth}{@{\extracolsep{\fill}}l }
\hline
Hsa.462 (Human serine kinase mRNA)\\
Hsa.549 (Transcription factor IIIA)\\
Hsa.601 (Human aspartyl-tRNA synthetase alpha-2 subunit mRNA)\\
Hsa.627 (Human monocyte-derived neutrophil-activating protein (MONAP) mRNA)\\
Hsa.773 (Macrophage migration inhibitory factor (Human))\\
Hsa.821 (Human hmgI mRNA for high mobility group protein Y)\\
Hsa.831 (Mitochondrial matrix protein P1 precursor (Human))\\
Hsa.957 (Human nucleolar protein (B23) mRNA)\\
Hsa.1832 (Myosin regulatory light chain 2, smooth muscle isoform (Human))\\
Hsa.2097 (Human vasoactive intestinal peptide (VIP) mRNA)\\
Hsa.2645 (H.sapiens ckshs2 mRNA for Cks1 protein homologue)\\
Hsa.2928 (H.sapiens mRNA for p cadherin)\\
Hsa.3016 (S-100P protein (Human))\\
Hsa.3306 (Human gene for heterogeneous nuclear ribonucleoprotein (hnRNP) core protein A1)\\
Hsa.3331 (Nucleoside diphosphate kinase A (Human))\\
Hsa.5971 (Human splicing factor SRp30c mRNA)\\
Hsa.6472 (Tubulin beta chain (Haliotis discus))\\
Hsa.6814 (Collagen alpha 2(XI) chain (Homo sapiens))\\
Hsa.8125 (Human)\\
Hsa.8147 (Human desmin gene)\\ 
Hsa.36689 (H.sapiens mRNA for GCAP-II/uroguanylin precursor)\\
Hsa.36952 (Complement factor D precursor (Homo sapiens))\\
Hsa.37937 (Myosyn heavy chain, nonmuscle (Gallus gallus) \\ 
\hline
\end{tabular*}}
\label{tab:colonvariables}
\end{table}

\end{document}